# Scalable on-chip multiplexing of silicon single and double quantum dots

Heorhii Bohuslavskyi[#], Alberto Ronzani, Joel Hätinen, Arto Rantala, Andrey Shchepetov, Panu Koppinen, Janne S. Lehtinen[#], and Mika Prunnila[#],

*VTT Technical Research Centre of Finland, Tietotie 3, 02150 Espoo, Finland*

*[#]Corresponding authors: heorhii.bohuslavskyi@vtt.fi, mika.prunnila@vtt.fi, and janne.lehtinen@semiqon.tech*

**Owing to the maturity of complementary metal oxide semiconductor (CMOS) microelectronics, qubits realized with spins in silicon quantum dots (QDs) are considered among the most promising technologies for building scalable quantum computers. For this goal, ultra-low-power on-chip cryogenic CMOS (cryo-CMOS) electronics for control, read-out, and interfacing of the qubits is an important milestone. We report on-chip interfacing of tunable electron and hole QDs by a 64-channel cryo-CMOS multiplexer with less-than-detectable static power dissipation. We analyze charge noise and measure state-of-the-art addition energies and gate lever arm parameters in the QDs. We correlate low noise in QDs and sharp turn-on characteristics in cryogenic transistors, both fabricated with the same gate stack. Finally, we demonstrate that our hybrid quantum-CMOS technology provides a route to scalable interfacing of a large number of QD devices, enabling, for example, variability analysis and QD qubit geometry optimization, which are prerequisites for building large-scale silicon-based quantum computers.**

## Introduction

To fully unlock the potential of quantum computers capable of performing certain computational tasks that are unfeasible with classical supercomputers **[1]**, millions of physical quantum bits (qubits) might be required **[2-4]**. Solid-state qubits based on electron and hole spins in semiconductor quantum dots (QDs) **[5-7]** are considered to provide one of the most scalable quantum computing platforms **[8]**. Historically, the first semiconductor spin qubits were demonstrated in GaAs in 2005 **[9]**. More recently, the progress in manufacturing commercial silicon transistors and CMOS circuitry - accumulated since the 1960s - has been harnessed for the fabrication of silicon QD spin qubit devices **[10-12]**, which culminated in the recent demonstration of an operational linear array of six electron spin qubits **[13]**. Similarly, a two-dimensional array of hole spin qubits in germanium has been recently reported **[14]**. Notably, the demonstration of the coherent control of spins in silicon QDs hosted in advanced complementary metal oxide semiconductor (CMOS) silicon field-effect transistors (FETs) **[10,11]** does not directly solve the challenge of the spin-based quantum information processing being extremely sensitive to background charge and (nuclear) spin environment **[8]**. In this respect, standard Si component processing can also accommodate isotopically purified silicon ($^{28}$Si), which leads to a reduction of the nuclear-spin-originated noise and, thereby, higher qubit fidelity. Si-based QD spin qubits can be realized using planar or nanowire (FinFET) QD geometries **[10-13,15,16]**. Furthermore, given the prospects of direct integration with on-chip classical cryogenic electronics used to initialize, drive, and read out qubits as well as perform quantum error correction **[17-20]**, silicon-based platforms have become one of the most studied routes for building large-scale quantum computers **[19]**.

For large-scale silicon quantum computers, extending the qubit operation to "hot" temperatures (above 1K) **[18]** is an important milestone toward monolithic integration between the physical spin qubit layer and auxiliary control and read-out electronics **[11,21,22]**. Moreover, large-scale analysis of mesoscopic spin qubit devices is required to engineer better qubits. In this respect, cryogenic signal multiplexing is considered extremely important for large-throughput characterization and, at the same time, an interfacing layer of quantum processors **[23-28]**. Therefore, low-power cryogenic

CMOS (cryo-CMOS) on-chip multiplexing quantum dot-based qubits is another crucial challenge to be solved to build large-scale silicon quantum computers **[17-19]**.

In this article, we report on the scalable interfacing of an array of electron and hole quantum dots with a monolithically integrated on-chip, ultra-low-power 64-channel cryo-CMOS multiplexer (MUX). Using a custom silicon-on-insulator CMOS fabrication process with the all-silicon gate stack, we demonstrate tunable electron and hole double quantum dots with large addition energies and gate lever arm parameters, measure low charge noise at 5.6 K of quantum dots connected to the access transistor switches integrated with cryo-CMOS multiplexer, and demonstrate scalable characterization of several dozens of quantum dot devices measured in the same cooldown. We demonstrate that our cryogenic multiplexing based on a digital decoder and analog switches is quasi-dissipationless from 5.6 K down to 300 mK and, thus, holds excellent potential for very large-scale characterization of silicon quantum dots and spin qubits.

# Results and Discussion

## Multiplexer, device selectivity, and cryo-CMOS

Figures 1 (a)-(f) show micrographs of the chip with monolithically integrated cryogenic CMOS MUX and quantum dot devices. An optical micrograph of the 64-channel multiplexer chip is shown in Fig. 1(a). A block of cryogenic conventional logic featuring standard CMOS NOT logic gates is shown in Fig. 1(b), and three parallel double quantum dot (DQD) devices are shown in Fig. 1(c). A cross-section of the metal oxide semiconductor field-effect transistor (MOSFET) which is used in the cryo-CMOS circuit is shown in Fig. 1(d). MUX-quantum dot components were realized on silicon-on-insulator wafers with a custom fabrication process involving an undoped channel, n++ doped Poly-Si (degenerately doped with phosphorus polysilicon)/$SiO_2$/Si in the front-end-of-line, and TiW + Al for the backend metallization. The tilted scanning electron microscopy (SEM) micrograph of a DQD device taken after the final passivation step of the back-end-of-line process is shown in Fig. 1 (e). The same fabrication process was used for the ambipolar double quantum dot devices recently reported in Ref. **[29]**. Here, the gate length ($L_g$) and gate pitch for all the quantum dot devices is 50 nm, and the SOI channel thickness ($W$) is 70 nm.

The cryogenic MUX is composed of a 6-to-64 decoder made of standard CMOS logic gates, and analog switches using a transmission gate design that features a pair of electron n- and hole p- MOSFETs. An optical micrograph of an electron double quantum dot (shown in Fig. 1(e)) which is connected to the analog switches is shown in Fig. 1 (f). The inset shows a circuit topology of the analog switch used to connect to a selected device. A simplified schematic of the MUX is shown in Fig. 1(g). The MUX logic part features a supply power line $V_{DD}$, ground contact (GND), and the 6 address line voltages, where A0(A5) corresponds to the lowest (highest) bit of the address bus. Overall, the MUX architecture can be split into 4 main blocks (see Fig. 2). These are: a data input/buffer with 6 address line inputs (A[0] – A[5]); 12 address lines connecting the address line voltages to the decoder; a digital decoder based on the combination of (three inputs) NAND3, NOR, and NOT logic gates (where NAND, NOR, and NOT correspond to the standard Boolean functions) repeated 64 times; 128 analog switches (combination of n- and p-MOSFET transistors); and the devices under test (from D[0] to D[63]). Source and drain terminals are connected through the MUX, and the 5 shared gate lines are directly routed to all the devices. A small DC voltage of a few mV ($V_{QD}$) is applied to the device bias input port, and device current output line is connected to a room-temperature transimpedance amplifier and a digital multimeter. Thus, to select one of the devices under test among 24 single electron and hole QDs, 24 electron and hole DQDs, and 16 test nanowires, a combination of the A[0] – A[5] voltages is supplied to the decoder while having $V_{DD}$ applied. As an illustration, to select the second device D[1], $V_{add}$ = {A[5] = "0", A[4] = "0", A[3] = "0", A[2] = "0", A[1] = "0", A[0] = "1"} is applied. Here, "0" and "1" are the logic zero and one. Then, for example, the electron DQD dev#1 (D[42]) introduced in Fig. 3 was selected by applying $V_{add}$ = {"1",

"0", "1", "0", "1", "0"} (or {101010} for short), see also the entire look-up table in Supplementary Table S1. The spacing between different devices in the MUX is 50 μm.

The role of $V_{\mathrm{add}}$ can be understood as follows: it corresponds to the direct ($V_{\mathrm{add}}$) and inverted ($\bar{V}_{\mathrm{add}}$) gate voltages applied to the n- and p-MOSFET of the analog switch, to have the switch transistor turned on into a low-resistive state. By measuring the source-drain current ($I_{\mathrm{DS}}$) as a function of gate voltage ($V_{\mathrm{GS}}$) of individual transistors, we estimated the pair of analog switches to add approximately a few kΩ of resistance in series with the MΩ-impedance QD device. The quantum dot experimental data presented in this article were acquired using $V_{\mathrm{add}}$ = $V_{\mathrm{DD}}$ = 1.5 V. Additionally, we show that using static substrate biasing (see Supplementary Figs. 1-2), $V_{\mathrm{DD}}$ could be lowered down to 0.8 V. We envision lowering $V_{\mathrm{DD}}$ even further by having two global independent back gates for n- and p-MOSFETs, as it was done in commercial low-power silicon-on-insulator technologies characterized at low temperatures **[30]**.

All 64 devices share the five chip gate terminals to drive the voltages applied to two gate-2 $G_{a1}$ & $G_{a2}$, two gate-1 $G_{b1}$ & $G_{b2}$, and one accumulation gate $G_{\mathrm{chan}}$, which can be routed at gate-1 or gate-2 level. Depending on the architecture of the selected device (see Supplementary Table S1 and Fig. 1(g)), voltages applied to $G_{a1(2)}$ and $G_{b1(2)}$ ($G_{b1(2)}$ and $G_{a1(2)}$) can correspond to $V_{\mathrm{pl1(2)}}$ and $V_{\mathrm{bar1(2)}}$ ($V_{\mathrm{bar1(2)}}$ and $V_{\mathrm{pl1(2)}}$). The difference between the two gate layers (for simplicity, referred later in the text as Gate-1 for the thinner and Gate-2 for the thicker gate oxide) is the gate oxide thickness: 20 nm and 55 nm, respectively.

The device selectivity using D[0] and D[1] for illustration, which are n- and p-type gated nanowires, is demonstrated in Supplementary Fig. 1. There, to confirm that the MUX decoder and switches function as designed, the field-effect transistor characteristic of devices D[0] and D[1] are shown, featuring the accumulation of electrons and holes in the nanowires D[0] and D[1], respectively. The cryogenic CMOS logic leakage current was found to be less than 1 pA (the noise floor of the used setup), which corresponds to the sub-pW cryogenic CMOS power dissipation level. We confirmed the static power dissipation of < 1 pW and correct device selectivity for all 64 devices measured in two different 64-channel MUXes at 5.6 K. Quantum transport measurements of coarse-resolution stability diagrams for 32 electron and hole, single and double quantum dots, and field-effect transistor measurements of 16 test nanowires from two different MUX chips (#1 and #2) are shown in Supplementary Figures 8-16.

The MOSFET characteristics $I_{\mathrm{DS}}(V_{\mathrm{GS}})$ of individual n- and p-MOSFETs measured at 5.6 K are shown in Supplementary Fig. 2. Since the same n++ Poly-Si was used in the gate stack of both n- and p-type transistors, the threshold voltages are not symmetric. Upon applying global back-biasing with $V_{\mathrm{BACK}} = -10$ V, almost symmetric cryogenic threshold voltages of n- and p-MOSFETs are demonstrated. The subthreshold swing $SS = \partial V_{\mathrm{GS}}/\partial \log(I_{\mathrm{DS}})$ as a function of source-drain current for n- and p-MOSFETs and quantum dot devices configured as MOSFETs in the linear source-drain bias regime is shown in Supplementary Fig. 3. The temperature dependence of $SS$ is given by $SS(T) = m\, ln10\; k_B T/e$, where $T$ is the temperature, $k_B$ is the Boltzmann constant, $e$ is the elementary charge, and $m$ is the technological parameter related to the fabrication process, which mainly results from a finite density of interface traps and disorder at the channel/gate oxide interface **[31]**. The lowest possible value of $SS$ in conventional MOSFETs at 300 K is 60 mV/dec, which corresponds to $m$ = 1. In our case, we obtain $m$ = 1.33 based on $SS$(300 K) = 80 mV/dec measured for n- and p-MOSFETs. Taking into account the band-broadening effects on the cryogenic subthreshold swing in MOSFETs, which result in the deviation from the linear temperature scaling of $SS$ below the saturation temperature **[32]**, our measured cryogenic subthreshold swing as low as 4 mV/dec in the weak inversion regime (characterized at 5.6 K) indicates a low-disorder environment of the MOSFET conductive channel. The 4 mV/dec value corresponds to approximately 1.5 meV band tails below the edges of conductance ($E_{\mathrm{c}}$) and valence ($E_{\mathrm{v}}$) bands. Based on the measured $SS$(5.6 K), we estimate the critical temperature below which the $SS$ is saturated to be $T_{\mathrm{C}} \sim 15$ K. Our minimal value of $SS$ and the critical temperature obtained for our gate stack with Poly-Si gates compare favorably with a commercial fully-depleted silicon-on-insulator technology based on a high-k metal gate process ($SS$ = 7−8 mV/dec and $T_C$ = 35 K, see Ref. **[32]**). Thus, our small cryogenic $SS$ values are very promising for building ultra-low-power cryogenic CMOS transistors (due to very sharp turning on of $I_{DS}(V_{GS})$ curves) and making low-disorder silicon quantum-dot-based spin qubits (as discussed latter) using the same platform.

Now, let us consider the cryogenic operation of the inverter as the basic CMOS logic gate (see Supplementary Fig 4). We measured the single inverter at both 2.5 K and 300 mK, and an almost identical operation for cryo-CMOS logic at both temperatures was found. Moreover, we confirmed the advantageous effect of negative back-gate biasing. The MOSFET threshold voltage tunability with back-biasing is shown in Supplementary Fig. 2. The latter makes the $V_{TH}$ of p- and n-MOSFETs more symmetrical, and the transistors are thus better matched. However, we show that our CMOS logic is also fully operational in the absence of back-gate biasing. Finally, we illustrate the potential of our cryo-CMOS technology for analog circuitry applications by analyzing the CMOS inverter's voltage gain. Using the negative back-gate biasing with $V_{BACK} = -10$ V, we demonstrated close matching of n- and p-MOSFETs despite using n++ Poly-gate for both transistors. The measured voltage gain of a single inverter reaches of $\sim 25-30$ at 2.5 K and 300 mK, which is comparable to the performance of commercial advanced CMOS circuitry **[33]**.

## Low-noise electron and hole double quantum dots measured with a MUX

Now that we have introduced and described in-depth the core results of this work – low-power cryo-CMOS multiplexing technology for on-chip measurements of various quantum mesoscopic devices such as quantum dots – we focus on several proof-of-concept experiments related to isolation and fine-tuning of selected electron and hole QDs embedded in the cryo-CMOS MUX. Even though the present cryogenic MUX chip featured 64 embedded quantum dot devices, only a few designs could be efficiently explored for the isolation and tuning of similar single and double quantum dots. However, as discussed later, we verified that even for the non-optimal from the point of view of tunability and gate-coupling strength quantum dot devices, the yield of working ohmic contacts and gates is high. Although most of the QD characterization data was acquired at 5.6 K due to the setup limitations, we also show that the same cryo-CMOS technology is operational down to 300 mK.

The schematic cross-section of a double quantum dot DQD device (inside the MUX) is shown in Fig. 3(a). The electron and hole DQDs reported in this Article did not feature dedicated access tunnel barriers that would have allowed a higher degree of control for tuning DQDs. Therefore, we focused on the measurements of only two electron and one hole DQD devices. To facilitate the electron DQD formation and make the access tunnel barriers connecting QDs to the reservoirs more opaque, back-gate biasing with $V_{BACK} = -10$ V was used to characterize two nominally identical electron DQDs. The back-gate is represented by standard slightly p-type $\sim 10^{15}\text{cm}^{-3}$ Si substrate beneath the buried oxide. Although the substrate was freezing out at low temperatures, we were able to operate it as a static gate with short settling times at relatively high temperatures of 5.6 K. We found almost the same charge noise, lever arm, and charging energy at $V_{BACK}$ = 0 V and −10 V applied through $t_{BOX}$ = 400 nm, which corresponds to very weak back-gating, as verified for QD2 of dev#1 described in Fig. 3 ($V_{BACK} = -10$ V). See Supplementary Fig. 17 for the data on the same device measured at $V_{BACK}$ = 0 V. In Fig. 3 (b,c), the DQD stability diagrams measured at 5.6 K obtained by sweeping two plunger (gate-1) gate voltages with $V_{pl1}$ and $V_{pl2}$ while fixing the inner barrier (gate-2) gate voltage $V_{bar1}$ are shown. The quantum dot source-drain bias was fixed to $V_{QD}$ = 2.2 mV and 1.8 mV and barrier gate voltage was set to $V_{bar1}$ = 3 V and 3.4 V for dev#1 (D[42]) and dev#2 (D[58]), respectively. The same accumulation gate voltage $V_{chan}$ = 3 V was used for both devices. A typical honeycomb pattern – the signature of two coupled quantum dots **[34]** – was obtained for dev#1 and dev#2 by measuring source-drain current $I_{QD}$ through the device as a function of two plunger gate voltages ($V_{pl1}$,$V_{pl2}$). The evolution of two weakly coupled QDs into strongly coupled, almost merged quantum dots as a function of $V_{bar1}$ is shown in Fig. 3(d).

Next, we focus on the Coulomb diamond measurements for which weakly coupled DQDs were configured with $V_{\mathrm{barl}}$ = 2.1 V and −1.5 V for dev#1 and #2, respectively. To probe Coulomb diamonds of QD1 (QD2), we filled the adjacent QD2 (QD1) with a dozen electrons (estimated to be approximately 10 – 20 electrons for dev#1 and 20 – 30 electrons for dev#2) and used the latter as an extended electron reservoir, see Fig. 3(e-f). While the current QD devices did not feature a charge sensor to ensure that the 1st electron was detected, judging by the gradually decreasing addition energy $E_{\mathrm{add}}$ (energy required to load an additional electron onto a QD, see **[34]**) as the QD is filled up, and no interruption of the opening edges of the first diamond, we assume that the few-electron regime for the isolated single quantum dots in DQD devices has been reached. It should be noted that the similar addition energy $E_{\mathrm{add}}$ for the first detected electrons of ~ 15-20 meV was measured in dev#1 and dev#2, and similar plunger and barrier voltages ($V_{\mathrm{barl}}$ = 2.3 V and $V_{\mathrm{barl}}$ = 2.1 V for Figs. 3 (e) and (f), respectively) were used to tune up both devices. We also tested that upon applying source-drain bias larger than 20 mV, the QD devices with similar geometry as the DQDs described above become field-effect transistors (see Supplementary Fig. 3). This is another experimental result toward the conclusion that the few-electron regime was reached.

Charge noise in silicon spin qubits is one of the limiting factors for improving qubit performance **[10-12, 35-40]**. The spin qubit's charge noise background couples to the spin via spin-orbit coupling, but it can also affect spin coherence through other mechanisms **[41]**. Unlike *III–V* nanowire-based spin-orbit qubits **[42,43]**, spin-orbit coupling is weak for electrons in silicon, but its impact on the spin coherence of silicon QDs accumulated at the interface between Si and gate oxide is not negligible. The charge noise issue becomes even more important for hole Si spin qubits having intrinsically strong spin-orbit coupling **[10,11]**. Further exploration and modeling of the charge noise impact on the operation of spin qubits defined in $^{28}$Si metal-oxide-semiconductor (MOS) structures, focusing on the few-electron or hole regime where the qubits are typically operated, constitutes an important research topic toward fault-tolerant quantum computing (see the recent experimental demonstration of two-qubit gates in silicon at the threshold of quantum error correction in Refs. **[44-46])**, provided the charge noise is the main spin coherence limitation mechanism **[36]**.

At the same time, one of the important milestones for practical silicon quantum computing is the integration of silicon spin qubits together with on-chip cryo-CMOS auxiliary electronics used to initialize, operate, and read out qubits, and perform quantum error correction at >1 K, where the thermal budget is much more relaxed (few W at the 4 K plate), compared to the cooling power available in the standard commercial dilution refrigerators at mK-temperatures (few tens of μW). Following the development of the so-called hot spin qubits **[18,19]**, several demonstrations of electron and hole Si spin qubits above 1 K were reported **[11,21,22]**. However, the reported hot qubit fidelities above 1 K were lower as compared to the mK-operation. One of the identified reasons was the charge noise, increasing with temperature in silicon quantum dots depending, for example, on the uniformity of charge fluctuators' distribution near the QDs **[47]**. Thus, the hot quantum dot charge noise optimization is one of the main challenges for large-scale quantum computing based on hot qubits operated above 1 K.

A horizontal cut of the Coulomb diamond dataset (dev#1, QD1) from the left panel of Fig. 3(e) is shown in Fig. 4(a). It was acquired at $V_{\mathrm{QD}}$ = 1 mV. The absence of gate hysteresis upon sweeping up and down the plunger gate voltage is a characteristic of high-quality poly-Si gates. The numerical derivative of the QD current versus plunger voltage dependence (transconductance, $g_{\mathrm{m\text{-}pl1}} = \partial I_{\mathrm{QD}}/\partial V_{\mathrm{p11}}$) required to convert the measured low-frequency current noise $S_{\mathrm{I}}(f)$ to equivalent charge noise $S_{\mathrm{e}}(f)$ is shown in Fig. 4(b). Quantifying the low-frequency charge noise at 1 Hz on the flanks of a Coulomb peak where the absolute value of transconductance is maximized is a common metric to benchmark quantum dot charge noise **[12,37-40]**. Since the variations in charge noise are expected for different Coulomb peaks as local charge defects can be activated, to test that the disorder and charge fluctuators are uniformly distributed across the device, which should result in *1/f* dependence of charge noise vs frequency, we performed current noise measurements for the first three resolved Coulomb peaks.

In Fig. 4(c), the current noise power spectral density (PSD) $S_I(f)$ curves are shown for the $V_{pl1}$ set to the Coulomb blockade regime (background), on the top of the 1st Coulomb peak, and the left and right flanks of the first three Coulomb peaks. The current PSD curves on the flanks of several Coulomb peaks approximately follow the $1/f$ slope, pointing toward uniformly distributed charge traps. On the flanks of the Coulomb peaks (unlike the top of the Coulomb peak), due to the local extremums in transconductance $g_{m\text{-}pl1}$, quantum dot chemical potential fluctuations dominate the current noise **[38]**. The described charge noise measurement technique can be correctly applied when the variation of quantum dot potential is much smaller than the Coulomb peak width, which is the case for the datasets presented in this article.

The charge noise $S_e(f)$ dependence for the first three resolved Coulomb peaks is shown in Fig. 4(d). The $S_e(f)$ data for the 1st Coulomb peak yielded 27 μeV $Hz^{-0.5}$at 1 Hz for QD1 of dev#1. $S_e(f)$ was calculated as $\alpha^2 S_I(f) g_{m-pl1}^{-2}$ **[38]**, where transconductance is $g_{m\text{-}pl1} = dI_{QD}/dV_{pl1}$, and lever arm $\alpha$ (given by quantum dot capacitance normalized by total capacitance) is calculated from the slopes of Coulomb peaks **[38]**. To avoid overestimating $g_{m\text{-}pl1(2)}$, numerical smoothing was applied after numerical derivation using Savitzky-Golay filtering. The raw and filtered $g_{m\text{-}pl1(2)}$ data are shown in Supplementary Fig. 5. Then, as we fill up the quantum dot with the 2nd and 3rd detected electrons, the charge noise becomes smaller. The latter is expected due to the partial screening of the charge traps as more electrons are added to the quantum dot and a decrease in gate-control efficiency that translates into a lower lever arm parameter **[47,48]**. Using the same technique, we measured the charge noise for QD2 of dev#1 and QD1-2 of dev#2 at $V_{QD}$ = 1 mV, see Fig. 4(e-g). Across two different (nominally identical) double quantum dots, we obtained a reasonable charge noise variability of $S_e$(1 Hz) between 15 and 27 μeV $Hz^{-0.5}$ for the first resolved electron in 4 different quantum dots at 5.6 K.

In our MUX architecture, the main potential contribution of MUX-added low-frequency noise comes from the analog switches composed of a pair of n- and p-MOSFETs. These transistors are operated in the strong inversion regime ($V_{GS} \gg V_{TH}$, where $V_{TH}$ is the threshold voltage), resulting in added series resistance of ~kΩ for high, MΩ-impedance quantum dot devices. Note that quantum dot devices and MOSFET transistors are realized with the same gate stack and overall fabrication technology. Therefore, we expect the measured quantum dot charge noise to be dominated by the charge fluctuators in the vicinity of the small quantum dot area (small multi-gate transistor), while for the transistors with the area $W \times L_g \gg 10$ μm$^2$, the MOSFET charge fluctuators' contribution to the measured noise should be significantly reduced given that voltage power spectral density (proportional to $S_I$) scales as $1/(W \times L_g)$ **[49].**

Approximating the QD system as a parallel-plate capacitor, the QD dimensions were estimated from the measured QD capacitance ($C_{QD}$) using $r_{QD} = \sqrt{\frac{C_{QD} \times t_{SiO2}}{\varepsilon_0 \times \varepsilon_{SiO2}}}$, where $t_{SiO2}$ is the gate oxide thickness (20 nm), the dielectric constant of gate oxide material is $\varepsilon_{SiO2}$ = 3.9, $\varepsilon_0$ is the vacuum permittivity, and $C_{QD} = (\alpha \times e)\, E_{add}{}^{-1}$. To estimate $C_{QD}$, we used the addition energy and the lever arm of the first resolved electrons.

Let us now focus on the electron QD variability across dev#1 and dev#2, each featuring two QDs. The addition energy, lever arm, low-frequency charge noise, and QD radius for the first resolved electrons are shown in Fig. 5. The average addition energy of 18 meV (Fig 5(a)), lever arm α = 0.33 eV $V^{-1}$ (Fig 5(b)), low-frequency charge noise $S_e$(1 Hz) = 22 μeV $Hz^{-0.5}$ Fig 5(c)), and the circular QD radius of $r_{QD}$ = 41 nm were obtained (Fig 5(d)). We also measured a hole DQD (D[43]) at T = 5.6 K using $V_{BACK}$ = 0 V in the few-hole regime using the same double-quantum dot geometry as electron DQD devices #1 and #2, see Supplementary Fig. 6. For the hole DQD device we found $E_{add}$ = 19 mV, α = 0.23 eV $V^{-1}$, and $r_{QD}$ = 33.5 nm, comparable to the lithographically defined quantum dot area. We measured hole QD charge noise $S_e$(1 Hz) = 28 μeV $Hz^{-0.5}$ in the few-hole regime, also comparable to the noise measured for our electron quantum dots.

Assuming a linear scaling of charge noise with temperature, in terms of power spectral density $S_e$ $[eV^2 Hz^{-1}] \propto T$, as it was previously reported in silicon QDs **[37]**, we extrapolate the charge noise $S_e$ (1 Hz) for the first detected electrons (holes) to 3 (3.8) μeV $Hz^{-0.5}$ having $\alpha$ = 0.33 (0.23) eV $V^{-1}$, respectively, at 100 mK, which is a typical electron temperature in well-designed and filtered spin qubit dilution cryostats. Our estimated charge noise levels are similar to those reported for low-disorder metal-oxide-silicon double quantum dots fabricated in academic cleanroom with Poly-Si/ $SiO_2$/Si MOS stack that showed charge noise of $S_e$(1 Hz) = 3.4 μeV $Hz^{-0.5}$ at 300 mK ($\alpha$ = 0.067 eV $V^{-1}$) **[50]**, but with almost 5 times larger lever arm parameter. Previously, the state-of-the-art silicon electron quantum dots with $\alpha$ = 0.12 eV $V^{-1}$ were characterized in the few-electron regime between 0.1 K and 4 K, yielding $S_e$(1 Hz) of 2 and 12 μeV $Hz^{-0.5}$, respectively **[37]**. Another recent demonstration of silicon electron spin qubit quantum dot devices on 300 mm wafers with $S_e$(1 Hz) as low as 3.6 (0.6) μeV $Hz^{-0.5}$ with lever arm $\alpha = 0.3$ (0.1) eV $V^{-1}$ measured at mK temperatures that used Poly-Si/ $SiO_2$/Si MOS stack strongly supports our gate-stack approach **[51,52]**.

While even the optimized short-channel ($L_g$ < 100nm) transistors from commercial advanced CMOS technologies feature threshold voltage variability of the order of 10−100 mV, $V_{TH}$-variability becomes more pronounced at low temperatures where, for example, the thermally activated transport in the subthreshold region is greatly suppressed. Thus, the cryogenic $V_{TH}$-variability can be larger than at 300 K **[53]**. Since $V_{TH}$ is defined by the gate metal work function, channel doping, built-in electrostatic potential, short-channel effects, and other parameters that challenge circuit designers trying to design reliable cryogenic CMOS electronics **[17]**, the situation with the quantum dot $V_{TH}$ (i.e., gate voltages at which the first electrons or holes are added to a quantum dot) is also far from being trivial. However, electrostatically defined quantum dots, unlike donor-based quantum dots (e.g. **[46]**), are expected to have the $V_{TH}$ of the first electrons and holes relatively close to the $V_{TH}$ of transistors fabricated in the same process. It is not unexpected to have a quantum dot's threshold voltage variability of a few hundred mV **[12,54]**. In this work, we observe that both electron and hole QD's $V_{TH}$ of ~ 0.2…0.3 V and ~ −1.2…−1.3 V, respectively (see Fig. 3 and Supplementary Fig. 6), approximately follow n- and p-MOSFET threshold voltages (see Supplementary Fig. 2), indicating that the quantum dots were shaped upon the gate-tuned energy level approaching $E_c$ and $E_v$. This is further supported by the estimated electron and hole QD sizes being close to the lithographically defined QD area.

For the sake of compactness, most of the datasets in this work focus only on the 5.6 K temperature. However, to validate the quantum dot and cryo-CMOS operation at T < 1 K, we measured at 300 mK an 8-channel MUX fabricated on the same wafer. This test MUX chip had devices with single electron QD geometries. Both the ultra-low-power, quasi-dissipationless cryogenic CMOS logic functionality and confinement of quantum dots were observed, thus validating the developed hybrid quantum-dot-CMOS process. The 300-mK Coulomb data together with the single QD device layout are given in Supplementary Fig. 7.

## Large throughput cryogenic characterization

Although we report the in-depth analysis of only a few QD devices, we verified that the gates could be swept at rates of a few kHz, thus enabling fast large-scale acquisition of stability diagrams by applying saw-tooth pulses and performing buffered acquisition. The gate voltage sweeping rate was not limited by cryogenic CMOS circuitry where only capacitance from the IN & OUT analog switches impacts the measurement bandwidth, estimated to be > 1 MHz for the current realization of transmission gate switches (the frequency bandwidth can be extended beyond 100 MHz by replacing the current transmission gate switch design by an 1×1 $\mu m^2$ n-MOSFET connected to the device from both ends), but having the high-impedance device connected by ~ 100 cm-long coaxial cable to the room temperature transimpedance amplifier, which resulted in the RC damping.

In terms of large throughput cryogenic characterization, on-chip CMOS cryogenic multiplexing presented in this work is shown to bring a considerable speed-up in performing cryogenic transport measurements of multi-quantum dot devices. Other mesoscopic or quantum devices can also be embedded in the described cryogenic MUX. With the present MUX implementation, the following important QD parameters should be possible to assess at large scale: measuring low-frequency charge noise measurements in the single dots; probing the tunability of interdot coupling, as required for double quantum dots to enable fast exchange interaction; tuning dots within the same one- or two-dimensional arrays into the few-electron and hole regime using devices equipped with the charge sensors; assessing basic spin physics by verifying that Pauli Spin Blockade is reproducible; and doing statistical analysis of the valley splitting across nominally identical electron QDs, as a critical parameter to optimize electron spin qubits, among others. The radio-frequency operation of the next generation of cryogenic multiplexers should also enable short, snapshot-like QD measurements to extract basic characteristics from large ensembles of quantum dots and other cryogenic components. Note that fast cryogenic MUX switching operation will likely generate non-negligible dissipation and may require additional thermal management considerations to be investigated.

# Conclusions

We demonstrate a hybrid quantum-dot-CMOS circuit, where a quasi-dissipationless cryo-CMOS multiplexer is monolithically integrated with single and double, electron and hole quantum dots. These quantum dots have low electron and hole charge noise at 5.6 K, and future experiments will clarify the low-frequency charge noise performance below 1 K. We correlate the relatively low charge noise with very small cryogenic subthreshold swings of conventional n- and p-MOSFETs, and hence small disorder in the Si channel. Our results suggest that the CMOS process with doped Poly-Si/$SiO_2$/Si MOS stack, commonly used for transistor manufacturing until the early 2000s, is a very promising technology for spin qubits. We also demonstrated the cryogenic operation of ambipolar CMOS transistors in this Article. They were fabricated on the same wafer as the cryogenic MUX devices.

Given the absence of static power dissipation down to sub-1 K temperatures, the cryogenic MUXes studied in this Article can be utilized in such variability and reliability analyses that rely on the measurements of millions of nominally identical quantum dot devices. Following the recent progress in the computer-assisted automated characterization of quantum dot devices **[55-58]**, a large-scale realization of our MUXes with the application of machine learning auto-tuning and the algorithms for statistical analysis of basic quantum dot and qubit features (such as addition energy, lever arm parameter, charge noise, and spin blockade conditions) can become an important enabler in scalable silicon-quantum-dot-based quantum computing.

Integration of the hybrid quantum-dot-CMOS circuit technology (by monolithic or heterogeneous means) with other cryogenic devices and microsystems, such as, e.g., CMOS-compatible superconducting Josephson field-effect transistors **[59]**, can lead to extended functionalities and speed up the deployment of quantum technologies. For example, solid-state refrigerators utilizing Si-based micro-fabrication could cool the most temperature-sensitive parts of the integrated microsystem to sub-1 K temperatures **[60,61]** whilst others remain above 1 K, thus simplifying the overall system infrastructure. The read-out of semiconductor qubits can also benefit from integration. Here, we envision integration (e.g. by flip-chip bonding) of high-quality HEMTs (high-electron-mobility-transistors) with minimal parasitic capacitances to enable fast and flexible spin qubit read-out with MHz bandwidth **[62-64],** or external cryogenic transimpedance amplifiers **[65,66],** alleviating the need for the complex implementation of radio-frequency reflectometry read-out circuitry **[67,68]**.

Finally, it is important to note that it is not yet settled whether electron or hole spin qubits hold better potential for large-scale silicon quantum computing. Beyond conventional silicon quantum-dot qubit devices based on either electron or hole spins **[8,19]**, there is more flexibility in building circuits with ambipolar transistors and QDs, which can host holes or electrons depending on the gate polarization as it was demonstrated **[29,69-71]**, see also Supplementary Fig. 18. Hybrid ambipolar or separate n- and p-type quantum-dot devices interfaced with CMOS circuits, including ultra-low power cryogenic multiplexing, can enable more fair statistical benchmarking of the single hole and electron spin qubits.

# Methods

**Fabrication.** The devices were fabricated on 150 mm silicon-on-insulator (SOI) wafers with a customized CMOS process in VTT's Micronova cleanroom facilities. The process consisted of 8 ultra-violet (UV) and 3 e-beam lithography layers. The SOI layer (Si channel) was thinned down to 35 nm by thermal oxidation and oxide stripping and patterned to form the nanowires. A 20 nm thermal $SiO_2$ was grown to provide the insulator between the silicon nanowires and first gate layer. This step reduced the Si layer to its final thickness of 24 nm. The first and second polycrystalline silicon gate layers (gate-1 and gate-2 levels) have thicknesses 50 nm and 80 nm and were degenerately doped with low-energy phosphorous ion implantation. The 35 nm thick $SiO_2$ dielectric layer between the polysilicon gate layers was grown by low-pressure chemical vapor deposition (LPCVD). Openings through the deposited dielectrics were etched on the source/drain regions of the SOI and phosphorous (n-type) or boron (p-type) implantation was used to dope these regions. A 250 nm thick $SiO_2$ was deposited with LPCVD, and the wafers were heated to 950 °C to activate the dopants and anneal the dielectrics. Contact holes for all three layers were etched with subsequent dry and wet etching processes. Finally, the metallization layer consisting of 25 nm TiW and 250 nm Al was deposited and patterned, and the wafers were treated with a forming gas anneal passivation. Note that in Fig. 1(d), the topmost layer which was left un-colored corresponds to the platinum layer deposited during cross-sectional imaging with Focused Ion Beam Scanning Electron Microscopy (FIB-SEM).

**Design of cryo-CMOS.** First, n- and p-MOSFET transistors with different gate lengths and channel widths were characterized at room temperature from several test wafers. Next, using the transistor compact modeling based on room temperature data, Cadence simulations were performed to validate the cryogenic-temperature-aware CMOS logic and switch operation. Although the design did not account for the cryogenic transistor characteristics, it was anticipated that the threshold voltages would take more positive and negative values for n- and p-MOSFET transistors, respectively, and the off-current of transistors will zero due to the suppression of the thermionic current below the threshold voltage at low temperatures. Finally, the MUX chip was mounted onto a microcontroller board with PCB shielding (standard JLCC84 package). The MUXes embedding 64 and 96 devices were tested at room temperature using a digital oscilloscope (MS0-X 2024A) by pulsing address voltage lines with up to 50 Hz repetition rate. The selectivity of different devices at room temperature was thus confirmed.

**Cryogenic setups and instrumentation.** The 5.6 K measurements were performed in a cryo-free refrigerator (Optistat by Oxford instruments). The 5.6 K setup did not feature any low-frequency cryogenic filtering, but only the intrinsic low-pass cut-off of resistive coaxial cables (cut-off frequency $f_{cut-off}$ between 50 and 100 MHz) used for DC biasing. The DC voltages were supplied using commercial off-shelf digital-to-analog converters DAC (Keysight 34951) and auxiliary voltage output channels of Zurich Instrument's Lock-in (mlfi model). PSD noise measurement parameters were 916 Hz sampling rate, 16384 points, 10 averages, frequency bandwidth limited to 150 Hz due to the built-in low-pass filtering of the transimpedance amplifier. A home-made voltage divider by 100 and first-order low-pass filter with $f_{cut-off}$ = 211 Hz was used to apply source-drain bias. Gate and CMOS logic voltages were filtered with home-made low-pass filters with $f_{cut-off}$ < 1 kHz. Device current was measured with a commercial low-noise transimpedance amplifier FEMTO DDPCA-300 and a commercial digital multimeter. The current noise measurements were performed by feeding the output of the transimpedance amplifier into the lock-in voltage input followed by FFT processing using mlfi lock-in's built-in spectrum analyzer. The measurements at 2.5 K and 300 mK were performed using a cryo-free version of Oxford Instrument's Heliox with a base temperature of 300 mK. There, the same DAC and lock-in were used to provide DC voltages, and same transimpedance amplifier followed by a commercial digital multimeter was used for measuring current. The Heliox setup featured cryogenic low-pass filtering that resulted in $f_{cut-off}$ of few kHz. The ambipolar transistor was measured in a cryogenic probe station with a base temperature of 3.5 K using a parameter analyzer's source measure units (SMUs) with the noise floor of approximately 10 nA.

# Acknowledgements

We would like to thank the VTT's operators and process engineers for supporting the fabrication in the OtaNano and VTT Micronova cleanroom facilities. We gratefully acknowledge the financial support from the European Union's Horizon 2020 research and innovation programme under Grant Agreement Nos. 688539 (http://mosquito.eu),766853 (http://www.efined-h2020.eu/) and 824109 European Microkelvin Platform (EMP), the Academy of Finland project QuMOS (project numbers 288907 and 287768), ETHEC (No. 322580), and Center of Excellence program project (No 312294). We also acknowledge funding from Business Finland through Quantum Technology Industrial (QuTI) project no. 128291. H.B acknowledges support from the Academy of Finland through the individual postdoctoral fellowship, project CRYOPROC (No. 350325) and the funding from Horizon Europe project Qu-Pilot project (No. 101113983). A. Ron. acknowledges the support from the Academy of Finland through his personal fellowship grant, project SUPSI (No. 356542).

# Author contributions

A.Ron., J.S.L., A.S., P.K, and M.P. were responsible for the device fabrication process design and development and the final device fabrication. A.Ran. designed and simulated cryo-CMOS logic. A.Ran. and A.Ron. tested the multiplexer functionality at room temperature. A.Ron. and J.S.L. designed quantum dot devices. H.B. performed the measurements at 5.6 K and 3.5 K with inputs from J.S.L., M.P., A.Ron., A.Ran., and J.H. A.Ron. performed the measurements at 2.5 K and 300 mK with input from H.B. J.H. performed FIB/SEM analysis and took all micrographs. H.B. analyzed the data, made the figures, and wrote the manuscript with input from the rest of the authors. M.P. and P.K. conceived the idea and initiated the project. J.S.L. and M.P. supervised the project.

# Conflict of interest

The authors declare no conflicts of interest.

# Data Availability

The data that support the findings of this study and analysis scripts are available from the corresponding authors upon reasonable request.

# Code availability

The codes used for the data acquisition for this work are obtained from the open-source Python packages QCoDeS (https://github.com/QCoDeS/Qcodes). The codes used for data analysis were obtained from the standard Python libraries such as matplotlib (https://matplotlib.org), numpy (https://numpy.org), and scipy (https://scipy.org). The used codes are available from the corresponding authors upon reasonable request.

# Correspondence and requests for materials

Should be addressed to H.B., M.P, and J.S.L.

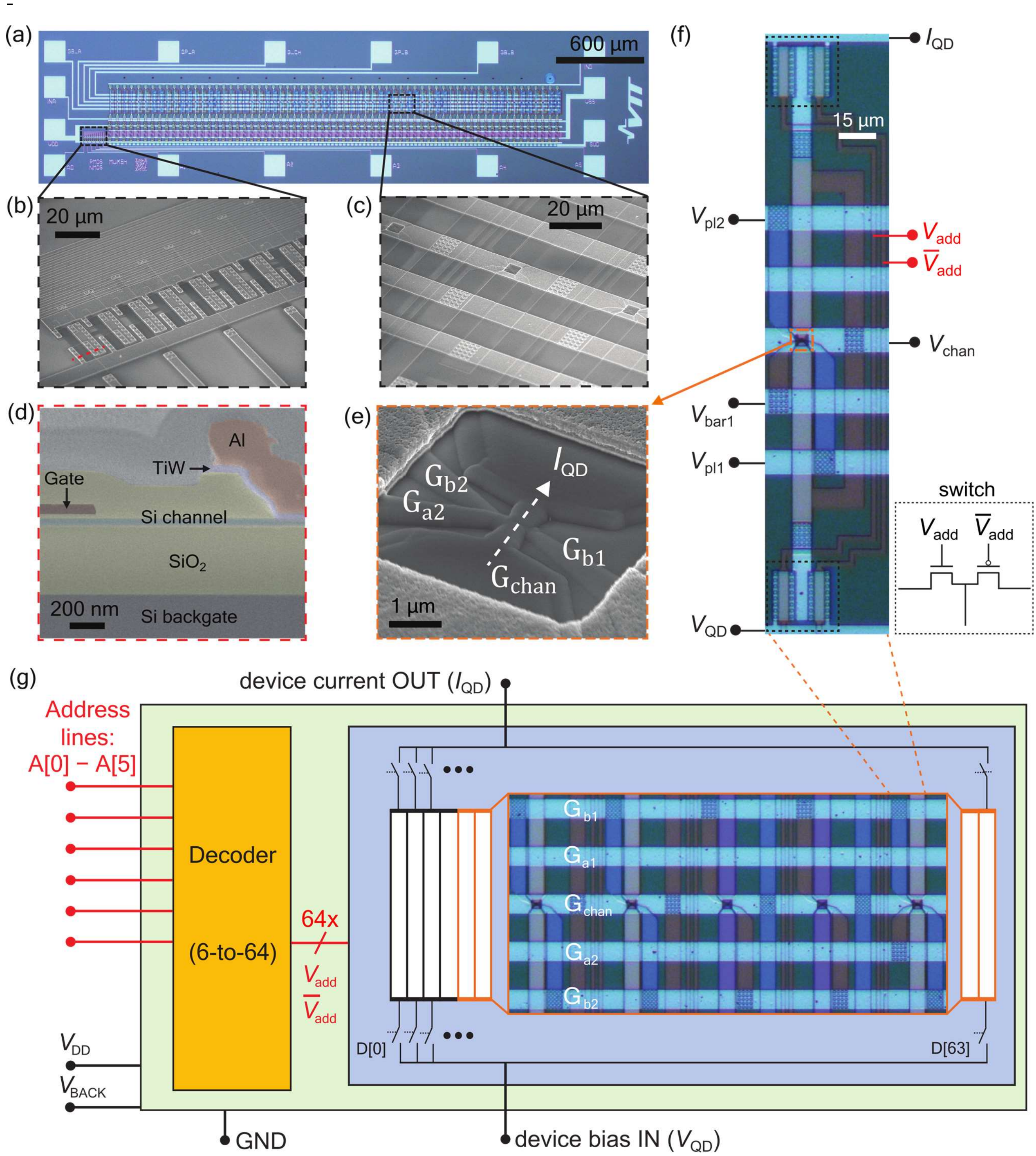

**Fig. 1 | Multiplexer, CMOS logic, and quantum dots. (a)** Optical micrograph of the 64-channel multiplexer chip. **(b)** Tilted scanning electron microscopy (SEM) image over cryogenic complementary metal oxide semiconductor (cryo-CMOS) logic. **(c)** Tilted SEM image over three double quantum dot devices. **(d)** Cross-section of a metal oxide semiconductor field effect transistor (MOSFET) used in the logic. The cut is taken along the dashed red line in (b). Color code used to highlight the key material stack layers is as follows: dark gray – Si (Si-bulk and silicon-on-insulator channel), yellow – $SiO_2$ (gate oxide and passivation), dark red – doped Poly-Si (gate metal), red and light blue – TiW and Al (metallization). **(e)** Tilted SEM over one of the double quantum dot (DQD) devices taken around the orange dashed rectangle in (f). The gates separated from the channel by 20 nm ($G_{b1}$, $G_{b2}$, and $G_{chan}$) and 55 nm ($G_{a2}$) of $SiO_2$ gate oxide are indicated. Quantum dot current $I_{QD}$ flowing between two electron reservoirs (Source and Drain) accumulated below

$G_{chan}$, through a DQD shaped with the voltages applied to $G_{b1}$, $G_{b2}$, and $G_{a2}$, is schematically shown. A schematic showing how the analog switch selection of a multiplexer-embedded device operates is given in **(f)**. The transmission gate switches (see the inset in (f)) are connected from both sides of the selected device. Address line voltages turn on the pair of analog switches that connect to the selected device with the direct and inverted multiplexer address line voltages $V_{add}$ and $\overline{V}_{add}$. Quantum dot source-drain bias $V_{QD}$ is applied between electron reservoirs that are accumulated with $V_{chan}$ voltage applied to $G_{chan}$. For the selected electron DQD device, plunger gate voltages ($V_{pl1}$, $V_{pl2}$) and barrier gate voltage ($V_{bar1}$) are applied to $G_{b1}$, $G_{b2}$, and $G_{a2}$, respectively. **(g)** Multiplexer schematic. It consists of 6 address lines (A[0] – A[5]) that are supplied to the decoder. The output of the decoder allows selecting one of the devices represented by black and orange rectangles that correspond to test nanowire and quantum dot devices respectively. The five gates ($G_{a1}$, $G_{a2}$, $G_{b1}$, $G_{b2}$, and $G_{chan}$) used to define and control D[0] – D[63] are highlighted in the optical micrograph. The cryogenic CMOS logic has a dedicated ground (GND) and is powered by supply voltage $V_{DD}$, and its operation can be adjusted with a global back-gate voltage $V_{BACK}$.

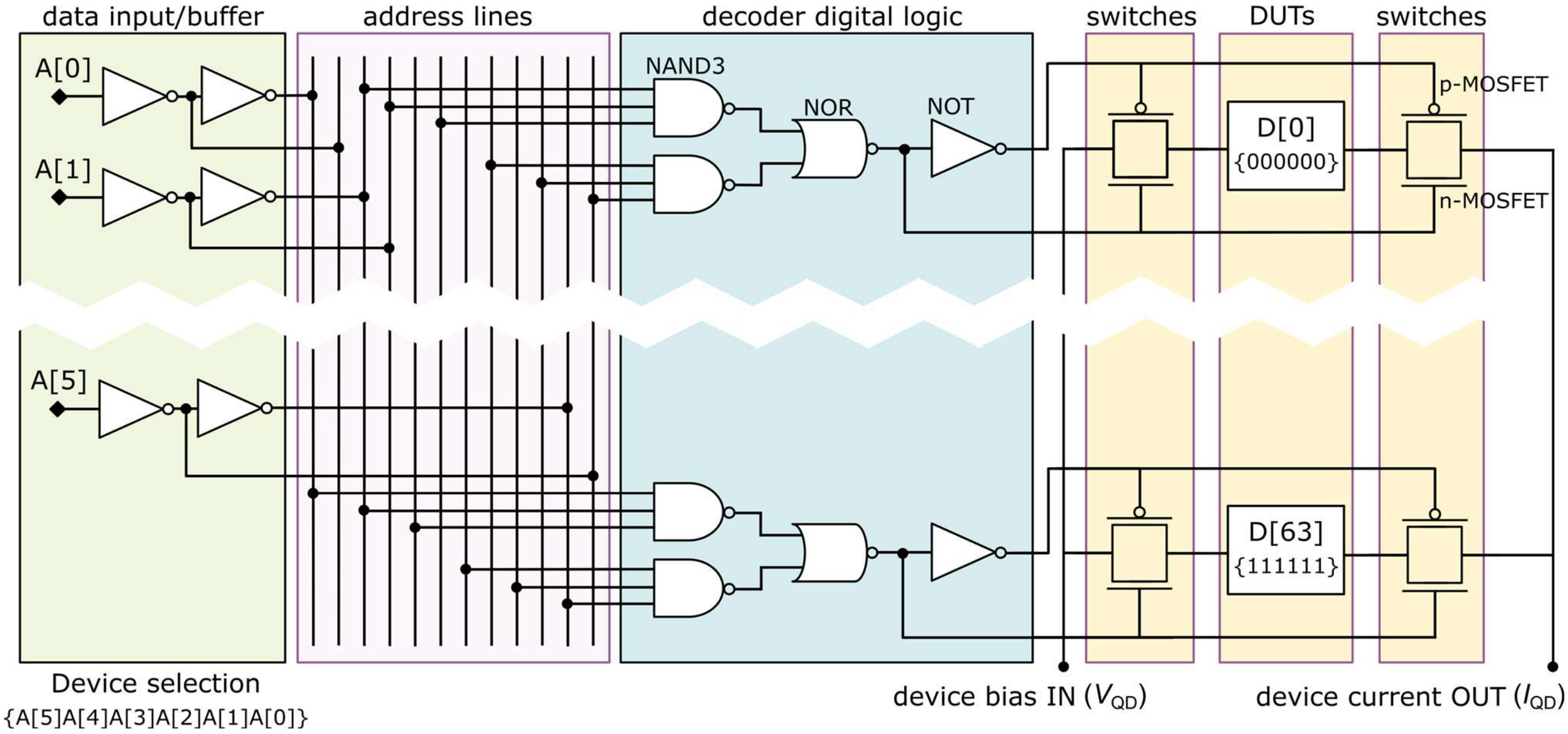

**Fig. 2 | Multiplexer architecture.** Main building blocks featuring data input block, address lines, decoder, and switches (composed of a pair of electron and hole metal oxide semiconductor field effect transistors, n- and p-MOSFETs) connected to Source/Drain contacts of 64 devices under test DUTs (Source/Drain multiplexing). The decoder digital logic block is composed of two NAND3, one NOR, and one NOT Boolean logic gate block, which is repeated 64 times inside the decoder. Complementary metal oxide semiconductor (CMOS) logic "1" and "0" applied to the six bits A[0] – A[5], {A[5]A[4]A[3]A[2]A[1]A[0]}, correspond to 1.5 V and 0 V, respectively. 64 devices from D[0] {000000} to D[63] {111111} can be independently selected and characterized in a single cooldown by applying quantum dot bias $V_{QD}$ and measuring quantum dot current $I_{QD}$.

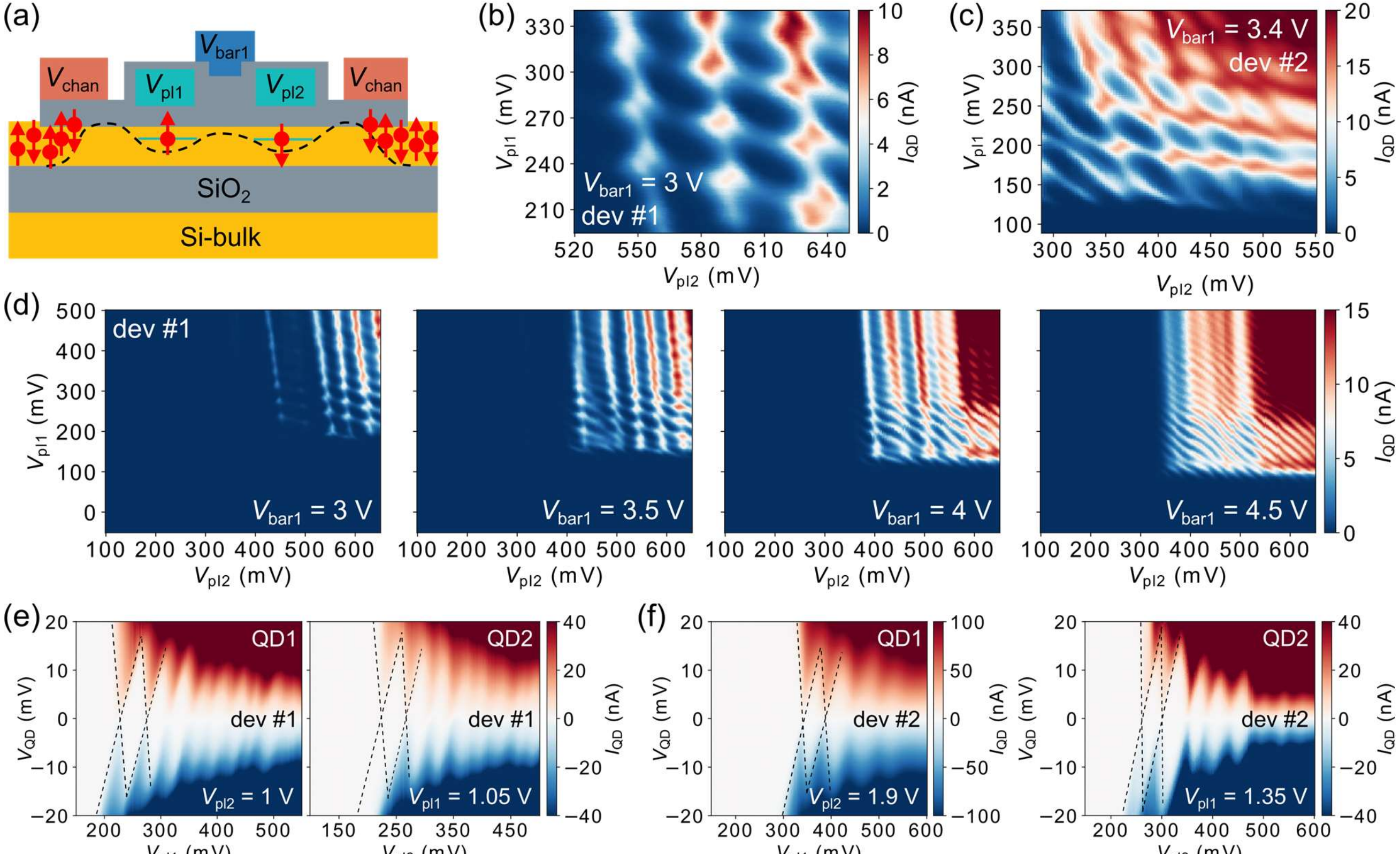

**Fig. 3 | Tunable few-electron double quantum dots**. Two nominally identical electron double quantum dot (DQD) devices #1 and #2 measured at back-gate voltage $V_{BACK} = -10$V and temperature $T = 5.6$ K. **(a)** A sketch showing the cross-section of the DQD device geometry studied in this figure with plunger (barrier) gates at gate-1 (gate-2) level. A tilted scanning electron microscopy image of such device is shown in Fig. 1(e). **(b-c)** Stability diagrams of coupled electron DQDs measured by sweeping plunger voltages $V_{pl1}$ and $V_{pl2}$ and measuring quantum dot current $I_{QD}$ for dev#1 and dev#2. **(d)** Tunability of the DQD (dev#1). By increasing the barrier voltage $V_{bar1}$ that controls the interdot tunnel coupling, from left to right: a weakly coupled DQD becomes a strongly coupled DQD. **(e-f)** Coulomb diamonds of quantum dot #1 controlled with $V_{pl1}$ (QD1) and quantum dot #2 controlled with $V_{pl2}$ (QD2) of dev#1 and #2 measured by filling up QD2 and QD1, respectively, with a dozen of electrons, and using the latter as an extended reservoir to probe QD1 and QD2. The black dashed lines indicate the first detected electrons.

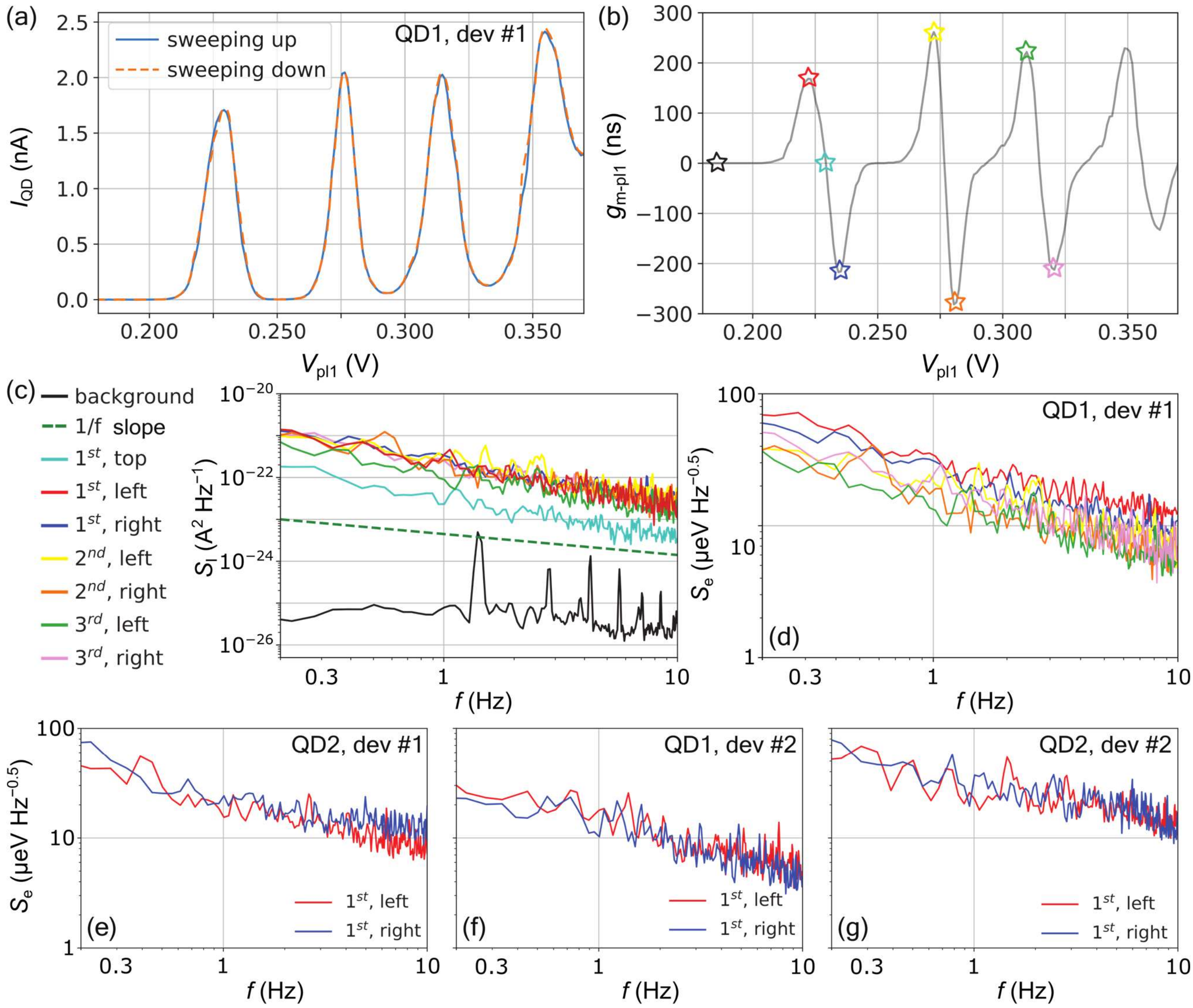

**Fig. 4 | Low-frequency charge noise in electron quantum dots. (a)** Coulomb peaks of quantum dot #1 (QD1) of dev#1 measured at quantum dot bias $V_{QD}$ = 1 mV and temperature $T$ = 5.6 K, which corresponds to a horizontal cut of the two-dimensional Coulomb map shown in Fig. 3(e). Note the absence of gate hysteresis demonstrated by sweeping up and down plunger gate voltage $V_{pl1}$. **(b)** Numerical derivative $dI_{QD}/dV_{pl1}$ of data in (a), where $I_{QD}$ is quantum dot current. The colored star markers are used to highlight the gate-voltage points at which noise measurements were performed. **(c)** Low-frequency current noise dependence on frequency $S_I(f)$ measured using the $V_{pl1}$ points indicated with the star markers of the same color in (b). The $S_I(f)$ curves are recorded at the $V_{pl1}$ points in the Coulomb blockade (background), on the top of the first Coulomb diamond (1st, top), and the right and left flanks of the first three detected Coulomb peaks (1st, 2nd, and 3rd left and right). Additionally, the $1/f$ frequency dependence trend ($1/f$ slope) is plotted. **(d)** Charge noise vs frequency $S_e(f)$ dependence of the first three detected electrons calculated from (c), measured at the flanks of Coulomb peaks. **(e-g)** The same charge noise experiment was performed for quantum dot #2 (QD2) of dev#1 and both QDs of dev#2.

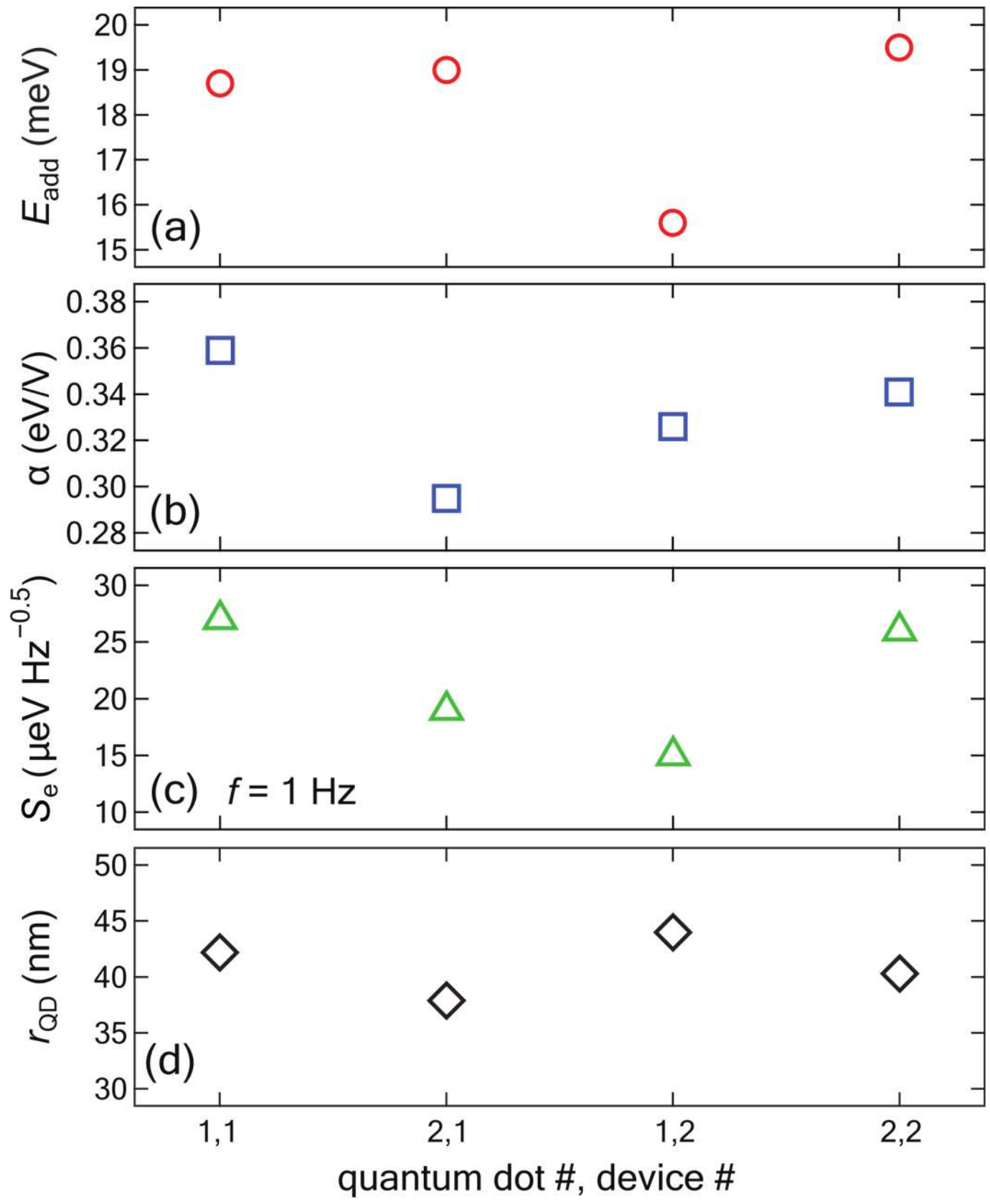

**Fig. 5 | Variability of electron quantum dots at 5.6 K. (a)** Addition energy ($E_{add}$), **(b)** lever arm parameter ($\alpha$), **(c)** low-frequency charge noise ($S_e$) extracted at frequency $f$ = 1 Hz, and **(d)** the estimated two-dimensional circular quantum dot (QD) radius ($r_{QD}$) are shown for four characterized quantum dots from electron double quantum dot (DQD) devices #1 and #2 (dev#1 and dev#2) described in the main text. The data are given for the first detected electrons. Note the low variability within the same DQD device and between two nominally identical DQD devices. The circular quantum dot radiuses are close to the lithographically defined quantum dot dimensions given by gate length multiplied by channel width $L_g \times W$ = 50 × 70 nm$^2$.

# Supplementary material for "*Scalable on-chip multiplexing of silicon single and double quantum dots*"

Heorhii Bohuslavskyi[#], Alberto Ronzani, Joel Hätinen, Arto Rantala, Andrey Shchepetov, Panu Koppinen, Janne S. Lehtinen[#], and Mika Prunnila[#]

*VTT Technical Research Centre of Finland, Tietotie 3, 02150 Espoo, Finland*

*[#]Corresponding authors: heorhii.bohuslavskyi@vtt.fi, mika.prunnila@vtt.fi, and janne@semiqon.tech*

**Supplementary Table S1 | Look-up table for all 64 devices embedded in the 64-channel cryogenic multiplexer.** The table includes device ID in short format (D[M], where N runs from 0 to 63) and long format ([XXXXXX], where five X's refer to CMOS logic binary "1" and "0" voltage). In the main text and all figures in the main text and supplementary material (except Supp. Fig. 1), address line voltage and supply voltage $V_{add} = V_{DD} = 1.5$ V and 0 V for supplying A[0] – A[5] address lines with logic "1" and "0" signal, respectively. Regarding the device type column, QD and DQD correspond to quantum dot and double quantum dot, and n- and p-type refers to electron and hole devices. The next column contains a list of gates and their level. The gate level gate-1 stands for 20 nm of $SiO_2$ gate oxide, and gate-2 corresponds to 50 nm of $SiO_2$ gate oxide. Also, the column specifies the gate voltages used to tune up and measure each device. The last column explains whether the data for each device is presented in the main text or supplementary material, or it was not measured or shown.

| Device ID (short) | Device ID (long) | Device type // idem devices | Gates (level) // Voltages | Measured devices |
| --- | --- | --- | --- | --- |
| D[0] | 000000 | gated nanowire, n-type // D[0], D[16], D[32], D[48] | $G_{chan}$ (gate-1) // $V_{chan}$ | Supp. Fig. 1 and 11 |
| D[1] | 000001 | gated nanowire, p-type // D[1], D[17], D[33], D[49] | $G_{chan}$ (gate-1) // $V_{chan}$ | Supp. Fig. 1 and 11 |
| D[2] | 000010 | gated nanowire, n-type // D[2], D[18], D[34], D[50] | $G_{chan}$ (gate-2) // $V_{chan}$ | Supp. Fig, 11 |
| D[3] | 000011 | gated nanowire, p-type // D[3], D[19], D[35], D[51] | $G_{chan}$ (gate-2) // $V_{chan}$ | Supp. Fig, 11 |
| D[4] | 000100 | single QD (design #1), n-type // D[4], D[20], D[36], D[52] | $G_{chan}$ (gate-2), $G_{b1,2}$ (gate-1) // $V_{chan}$, $V_{bar1}$, $V_{bar2}$ | Supp. Fig. 8 and 12 |
| D[5] | 000101 | single QD (design #1), p-type // D[5], D[21], D[37], D[53] | $G_{chan}$ (gate-2), $G_{b1,2}$ (gate-1) // $V_{chan}$, $V_{bar1}$, $V_{bar2}$ | Supp. Fig. 9 |
| D[6] | 000110 | single QD (design #2), n-type // D[6], D[22], D[38], D[54] | $G_{chan}$ (gate-1), $G_{b1,2}$ (gate-2) // $V_{chan}$, $V_{bar1}$, $V_{bar2}$ | Supp. Fig. 13 |
| D[7] | 000111 | single QD (design #2), p-type // D[7], D[23], D[39], D[55] | $G_{chan}$ (gate-1), $G_{b1,2}$ (gate-2) // $V_{chan}$, $V_{bar1}$, $V_{bar2}$ | Supp. Fig. 10 and 15 |
| D[8] | 001000 | single QD (design #3), n-type // D[8], D[24], D[40], D[56] | $G_{chan}$ and $G_{a1}$ (gate-1), $G_{b1,2}$ (gate-2) // $V_{chan}$, $V_{pl1}$, $V_{bar1}$, $V_{bar2}$ | *not shown (not all 4 devices were measured)* |
| D[9] | 001001 | single QD (design #3), p-type // D[9], D[25], D[41], D[57] | $G_{chan}$ and $G_{a1}$ (gate-1), $G_{b1,2}$ (gate-2) // $V_{chan}$, $V_{pl1}$, $V_{bar1}$, $V_{bar2}$ | *not shown (not all 4 devices were measured)* |
| D[10] | 001010 | DQD (design #1), n-type // D[10], D[26], D[42], D[58] | $G_{a1,2}$ (gate-1), $G_{chan}$ and $G_{b1}$ (gate-2) // $V_{chan}$, $V_{pl1}$, $V_{pl2}$, $V_{bar1}$ | Supp. Fig. 14 |
| D[11] | 001011 | DQD (design #1), p-type // D[11], D[27], D[43], D[59] | $G_{a1,2}$ (gate-1), $G_{chan}$ and $G_{b1}$ (gate-2) // $V_{chan}$, $V_{pl1}$, $V_{pl2}$, $V_{bar1}$ | Supp. Fig. 16 |
| D[12] | 001100 | DQD (design #2), n-type // D[12], D[28], D[44], D[60] | $G_{a1,2}$ and $G_{chan}$ (gate-1), $G_{b1,2}$ (gate-2) // $V_{chan}$, $V_{pl1}$, $V_{pl2}$, $V_{bar1}$, $V_{bar2}$ | *not measured* |
| D[13] | 001101 | DQD (design #2), p-type // D[13], D[29], D[45], D[61] | $G_{a1,2}$ and $G_{chan}$ (gate-1), $G_{b1,2}$ (gate-2) // $V_{chan}$, $V_{pl1}$, $V_{pl2}$, $V_{bar1}$, $V_{bar2}$ | *not measured* |
| D[14] | 001110 | DQD (design #3), n-type // D[14], D[30], D[46], D[62] | $G_{a1,2}$ and $G_{chan}$ (gate-2), $G_{b1,2}$ (gate-1) // $V_{chan}$, $V_{pl1}$, $V_{pl2}$, $V_{bar1}$, $V_{bar2}$ | *not measured* |
| D[15] | 001111 | DQD (design #3), p-type // D[15], D[31], D[47], D[63] | $G_{a1,2}$ and $G_{chan}$ (gate-2), $G_{b1,2}$ (gate-1) // $V_{chan}$, $V_{pl1}$, $V_{pl2}$, $V_{bar1}$, $V_{bar2}$ | *not measured* |
| D[16] | 010000 | gated nanowire, n-type // D[0], D[16], D[32], D[48] | $G_{chan}$ (gate-1) // $V_{chan}$ | Supp. Fig. 11 |
| D[17] | 010001 | gated nanowire, p-type // D[1], D[17], D[33], D[49] | $G_{chan}$ (gate-1) // $V_{chan}$ | Supp. Fig. 11 |
| D[18] | 010010 | gated nanowire, n-type // | $G_{chan}$ (gate-2) // | Supp. Fig. 11 |

| | | D[2], D[18], D[34], D[50] | $V_{chan}$ | |
|---|---|---|---|---|
| D[19] | 010011 | gated nanowire, p-type // D[3], D[19], D[35], D[51] | $G_{chan}$ (gate-2) // $V_{chan}$ | Supp. Fig. 11 |
| D[20] | 010100 | single QD (design #1), n-type // D[4], D[20], D[36], D[52] | $G_{chan}$ (gate-2), $G_{b1,2}$ (gate-1) // $V_{chan}$, $V_{bar1}$, $V_{bar2}$ | Supp. Fig. 8 and 12 |
| D[21] | 010101 | single QD (design #1), p-type // D[5], D[21], D[37], D[53] | $G_{chan}$ (gate-2), $G_{b1,2}$ (gate-1) // $V_{chan}$, $V_{bar1}$, $V_{bar2}$ | Supp. Fig. 9 |
| D[22] | 010110 | single QD (design #2), n-type // D[6], D[22], D[38], D[54] | $G_{chan}$ (gate-1), $G_{b1,2}$ (gate-2) // $V_{chan}$, $V_{bar1}$, $V_{bar2}$ | Supp. Fig. 13 |
| D[23] | 010111 | single QD (design #2), p-type // D[7], D[23], D[39], D[55] | $G_{chan}$ (gate-1), $G_{b1,2}$ (gate-2) // $V_{chan}$, $V_{bar1}$, $V_{bar2}$ | Supp. Fig. 10 and 15 |
| D[24] | 011000 | single QD (design #3), n-type // D[8], D[24], D[40], D[56] | $G_{chan}$ and $G_{a1}$ (gate-1), $G_{b1,2}$ (gate-2) // $V_{chan}$, $V_{pl1}$, $V_{bar1}$, $V_{bar2}$ | *not shown (not all 4 devices were measured)* |
| D[25] | 011001 | single QD (design #3), p-type // D[9], D[25], D[41], D[57] | $G_{chan}$ and $G_{a1}$ (gate-1), $G_{b1,2}$ (gate-2) // $V_{chan}$, $V_{pl1}$, $V_{bar1}$, $V_{bar2}$ | *not shown (not all 4 devices were measured)* |
| D[26] | 011010 | DQD (design #1), n-type // D[10], D[26], D[42], D[58] | $G_{a1,2}$ (gate-1), $G_{chan}$ and $G_{b1}$ (gate-2) // $V_{chan}$, $V_{pl1}$, $V_{pl2}$, $V_{bar1}$ | Supp. Fig. 14 |
| D[27] | 011011 | DQD (design #1), p-type // D[11], D[27], D[43], D[59] | $G_{a1,2}$ (gate-1), $G_{chan}$ and $G_{b1}$ (gate-2) // $V_{chan}$, $V_{pl1}$, $V_{pl2}$, $V_{bar1}$ | Supp. Fig. 16 |
| D[28] | 011100 | DQD (design #2), n-type // D[12], D[28], D[44], D[60] | $G_{a1,2}$ and $G_{chan}$ (gate-1), $G_{b1,2}$ (gate-2) // $V_{chan}$, $V_{pl1}$, $V_{pl2}$, $V_{bar1}$, $V_{bar2}$ | *not measured* |
| D[29] | 011101 | DQD (design #2), p-type // D[13], D[29], D[45], D[61] | $G_{a1,2}$ and $G_{chan}$ (gate-1), $G_{b1,2}$ (gate-2) // $V_{chan}$, $V_{pl1}$, $V_{pl2}$, $V_{bar1}$, $V_{bar2}$ | *not measured* |
| D[30] | 011110 | DQD (design #3), n-type // D[14], D[30], D[46], D[62] | $G_{a1,2}$ and $G_{chan}$ (gate-2), $G_{b1,2}$ (gate-1) // $V_{chan}$, $V_{pl1}$, $V_{pl2}$, $V_{bar1}$, $V_{bar2}$ | *not measured* |
| D[31] | 011111 | DQD (design #3), p-type // D[15], D[31], D[47], D[63] | $G_{a1,2}$ and $G_{chan}$ (gate-2), $G_{b1,2}$ (gate-1) // $V_{chan}$, $V_{pl1}$, $V_{pl2}$, $V_{bar1}$, $V_{bar2}$ | *not measured* |
| D[32] | 100000 | gated nanowire, n-type // D[0], D[16], D[32], D[48] | $G_{chan}$ (gate-1) // $V_{chan}$ | Supp. Fig. 11 |
| D[33] | 100001 | gated nanowire, p-type // D[1], D[17], D[33], D[49] | $G_{chan}$ (gate-1) // $V_{chan}$ | Supp. Fig. 11 |
| D[34] | 100010 | gated nanowire, n-type // D[2], D[18], D[34], D[50] | $G_{chan}$ (gate-2) // $V_{chan}$ | Supp. Fig. 11 |
| D[35] | 100011 | gated nanowire, p-type // D[3], D[19], D[35], D[51] | $G_{chan}$ (gate-2) // $V_{chan}$ | Supp. Fig. 11 |
| D[36] | 100100 | single QD (design #1), n-type // D[4], D[20], D[36], D[52] | $G_{chan}$ (gate-2), $G_{b1,2}$ (gate-1) // $V_{chan}$, $V_{bar1}$, $V_{bar2}$ | Supp. Fig. 8 and 12 |
| D[37] | 100101 | single QD (design #1), p-type // D[5], D[21], D[37], D[53] | $G_{chan}$ (gate-2), $G_{b1,2}$ (gate-1) // $V_{chan}$, $V_{bar1}$, $V_{bar2}$ | Supp. Fig. 9 |
| D[38] | 100110 | single QD (design #2), n-type // D[6], D[22], D[38], D[54] | $G_{chan}$ (gate-1), $G_{b1,2}$ (gate-2) // $V_{chan}$, $V_{bar1}$, $V_{bar2}$ | Supp. Fig. 13 |
| D[39] | 100111 | single QD (design #2), p-type // D[7], D[23], D[39], D[55] | $G_{chan}$ (gate-1), $G_{b1,2}$ (gate-2) // $V_{chan}$, $V_{bar1}$, $V_{bar2}$ | Supp. Fig. 10 and 15 |
| D[40] | 101000 | single QD (design #3), n-type // D[8], D[24], D[40], D[56] | $G_{chan}$ and $G_{a1}$ (gate-1), $G_{b1,2}$ (gate-2) // $V_{chan}$, $V_{pl1}$, $V_{bar1}$, $V_{bar2}$ | *not shown (not all 4 devices were measured)* |
| D[41] | 101001 | single QD (design #3), p-type // D[9], D[25], D[41], D[57] | $G_{chan}$ and $G_{a1}$ (gate-1), $G_{b1,2}$ (gate-2) // $V_{chan}$, $V_{pl1}$, $V_{bar1}$, $V_{bar2}$ | *not shown (not all 4 devices were measured)* |
| D[42] | 101010 | DQD (design #1), n-type // D[10], D[26], D[42], D[58] | $G_{a1,2}$ (gate-1), $G_{chan}$ and $G_{b1}$ (gate-2) // $V_{chan}$, $V_{pl1}$, $V_{pl2}$, $V_{bar1}$ | Fig. 1,3,4 and 5; Supp. Fig. 14 and 17 |
| D[43] | 101011 | DQD (design #1), p-type // D[11], D[27], D[43], D[59] | $G_{a1,2}$ (gate-1), $G_{chan}$ and $G_{b1}$ (gate-2) // $V_{chan}$, $V_{pl1}$, $V_{pl2}$, $V_{bar1}$ | Supp. Fig. 6 and 16 |
| D[44] | 101100 | DQD (design #2), n-type // D[12], D[28], D[44], D[60] | $G_{a1,2}$ and $G_{chan}$ (gate-1), $G_{b1,2}$ (gate-2) // $V_{chan}$, $V_{pl1}$, $V_{pl2}$, $V_{bar1}$, $V_{bar2}$ | *not measured* |
| D[45] | 101101 | DQD (design #2), p-type // D[13], D[29], D[45], D[61] | $G_{a1,2}$ and $G_{chan}$ (gate-1), $G_{b1,2}$ (gate-2) // $V_{chan}$, $V_{pl1}$, $V_{pl2}$, $V_{bar1}$, $V_{bar2}$ | *not measured* |
| D[46] | 101110 | DQD (design #3), n-type // D[14], D[30], D[46], D[62] | $G_{a1,2}$ and $G_{chan}$ (gate-2), $G_{b1,2}$ (gate-1) // $V_{chan}$, $V_{pl1}$, $V_{pl2}$, $V_{bar1}$, $V_{bar2}$ | *not measured* |
| D[47] | 101111 | DQD (design #3), p-type // D[15], D[31], D[47], D[63] | $G_{a1,2}$ and $G_{chan}$ (gate-2), $G_{b1,2}$ (gate-1) // $V_{chan}$, $V_{pl1}$, $V_{pl2}$, $V_{bar1}$, $V_{bar2}$ | *not measured* |
| D[48] | 110000 | gated nanowire, n-type // D[0], D[16], D[32], D[48] | $G_{chan}$ (gate-1) // $V_{chan}$ | Supp. Fig. 11 |
| D[49] | 110001 | gated nanowire, p-type // D[1], D[17], D[33], D[49] | $G_{chan}$ (gate-1) // $V_{chan}$ | Supp. Fig. 11 |
| D[50] | 110010 | gated nanowire, n-type // D[2], D[18], D[34], D[50] | $G_{chan}$ (gate-2) // $V_{chan}$ | Supp. Fig. 11 |
| D[51] | 110011 | gated nanowire, p-type // D[3], D[19], D[35], D[51] | $G_{chan}$ (gate-2) // $V_{chan}$ | Supp. Fig. 11 |
| D[52] | 110100 | single QD (design #1), n-type // D[4], D[20], D[36], D[52] | $G_{chan}$ (gate-2), $G_{b1,2}$ (gate-1) // $V_{chan}$, $V_{bar1}$, $V_{bar2}$ | Supp. Fig. 8 and 12 |
| D[53] | 110101 | single QD (design #1), p-type // D[5], D[21], D[37], D[53] | $G_{chan}$ (gate-2), $G_{b1,2}$ (gate-1) // $V_{chan}$, $V_{bar1}$, $V_{bar2}$ | Supp. Fig. 9 |
| D[54] | 110110 | single QD (design #2), n-type // D[6], D[22], D[38], D[54] | $G_{chan}$ (gate-1), $G_{b1,2}$ (gate-2) // $V_{chan}$, $V_{bar1}$, $V_{bar2}$ | Supp. Fig. 13 |
| D[55] | 110111 | single QD (design #2), p-type // D[7], D[23], D[39], D[55] | $G_{chan}$ (gate-1), $G_{b1,2}$ (gate-2) // $V_{chan}$, $V_{bar1}$, $V_{bar2}$ | Supp. Fig. 10 and 15 |

| D[56] | 111000 | single QD (design #3), n-type // D[8], D[24], D[40], D[56] | $G_{chan}$ and $G_{a1}$ (gate-1), $G_{b1,2}$ (gate-2) // $V_{chan}$, $V_{pl1}$, $V_{bar1}$, $V_{bar2}$ | Supp. Fig. 3 |
|---|---|---|---|---|
| D[57] | 111001 | single QD (design #3), p-type // D[9], D[25], D[41], D[57] | $G_{chan}$ and $G_{a1}$ (gate-1), $G_{b1,2}$ (gate-2) // $V_{chan}$, $V_{pl1}$, $V_{bar1}$, $V_{bar2}$ | Supp. Fig. 3 |
| D[58] | 111010 | DQD (design #1), n-type // D[10], D[26], D[42], D[58] | $G_{a1,2}$ (gate-1), $G_{chan}$ and $G_{b1}$ (gate-2) // $V_{chan}$, $V_{pl1}$, $V_{pl2}$, $V_{bar1}$ | Supp. Fig. 14 |
| D[59] | 111011 | DQD (design #1), p-type // D[11], D[27], D[43], D[59] | $G_{a1,2}$ (gate-1), $G_{chan}$ and $G_{b1}$ (gate-2) // $V_{chan}$, $V_{pl1}$, $V_{pl2}$, $V_{bar1}$ | Supp. Fig. 16 |
| D[60] | 111100 | DQD (design #2), n-type // D[12], D[28], D[44], D[60] | $G_{a1,2}$ and $G_{chan}$ (gate-1), $G_{b1,2}$ (gate-2) // $V_{chan}$, $V_{pl1}$, $V_{pl2}$, $V_{bar1}$, $V_{bar2}$ | *not measured* |
| D[61] | 111101 | DQD (design #2), p-type // D[13], D[29], D[45], D[61] | $G_{a1,2}$ and $G_{chan}$ (gate-1), $G_{b1,2}$ (gate-2) // $V_{chan}$, $V_{pl1}$, $V_{pl2}$, $V_{bar1}$, $V_{bar2}$ | *not measured* |
| D[62] | 111110 | DQD (design #3), n-type // D[14], D[30], D[46], D[62] | $G_{a1,2}$ and $G_{chan}$ (gate-2), $G_{b1,2}$ (gate-1) // $V_{chan}$, $V_{pl1}$, $V_{pl2}$, $V_{bar1}$, $V_{bar2}$ | *not measured* |
| D[63] | 111111 | DQD (design #3), p-type // D[15], D[31], D[47], D[63] | $G_{a1,2}$ and $G_{chan}$ (gate-2), $G_{b1,2}$ (gate-1) // $V_{chan}$, $V_{pl1}$, $V_{pl2}$, $V_{bar1}$, $V_{bar2}$ | *not measured* |

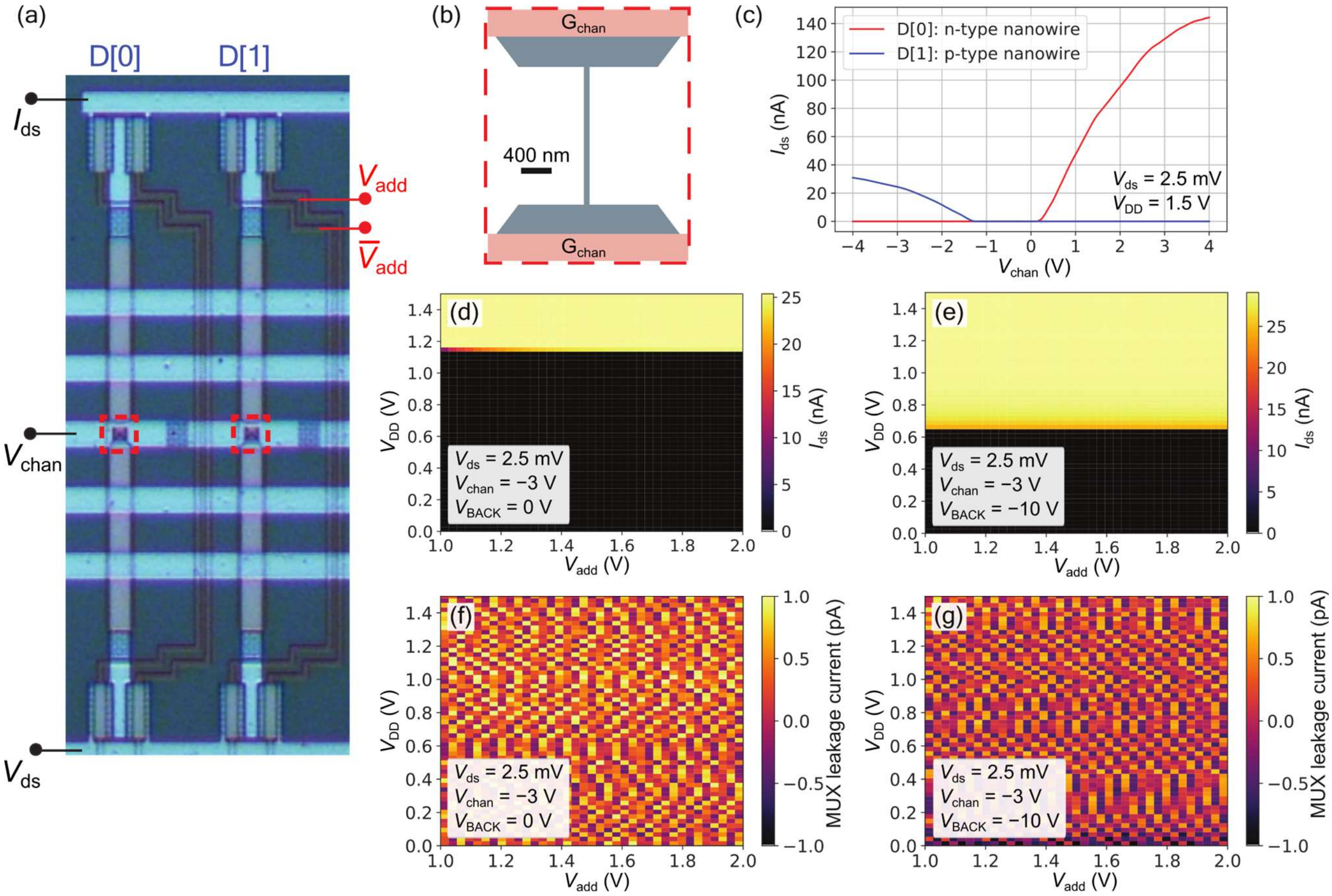

**Supplementary Fig. 1 | Quasi-dissipationless operation of cryo-CMOS multiplexer. (a)** An optical micrograph showing n- and p-type nanowires (D[0] and D[1] highlighted with red dashed squares) connected to device bias IN and device current OUT pairs of analog switches (see also Fig. 1(g) and Fig. 2 in the main text). The direct and inverted multiplexer address line voltages $V_{add}$ and $\overline{V}_{add}$ are also shown schematically. **(b)** A schematic of the nanowire geometry of devices D[0] and D[1]. Source-drain bias $V_{ds}$ is applied from one side (device bias IN) and source-drain current $I_{ds}$ is measured from another side (device current OUT). Electron and hole reservoirs for source and drain are induced by applying voltage $V_{chan}$ to the gate $G_{chan}$ at gate-2 level. **(c)** The device selectivity (depending on the applied address line voltages) is shown by measuring n-type nanowire (accumulation of electrons at positive $V_{chan}$) and p-type nanowire (accumulation of holes at negative $V_{chan}$) using $V_{ds}$ = 2.5 mV, supply voltage $V_{DD}$ = 1.5 V, and $V_{add}$ = 1.5 V. With $V_{add}$ =

{A[5] = 0 V, A[4] = 0 V, A[3] = 0 V, A[2] = 0 V, A[1] = 0 V, A[0] = 0 V}, D[0] or {000000} is selected, and $V_{add}$ = {0 V, 0 V, 0 V, 0 V, 0 V, X} connects D[1] or {000001}. X voltage value (logical "1") is typically above 1 V, and its absolute (and inverted) value such that, for example, 1.5 V (and −1.5 V) is chosen to bring electron and hole metal oxide semiconductor field effect transistors (n- and p-MOSFETs) along device bias IN and device current OUT analog switches. **(d)** The illustration of how the optimal $V_{DD}$ and $V_{add}$ are defined using D[1] (p-type nanowire) as an example. Global back-gate voltage $V_{BACK}$ = 0 V (applicable for cryo-CMOS transistors, quantum dot and nanowire devices at the same time) is used and negative $V_{chan}$ is applied to accumulate holes in the nanowire channel for D[1]. By measuring the current through D[1] as $V_{DD}$ and $V_{add}$ are swept, $I_{ds}$ starts to saturate at $V_{DD} \approx 1.2$ V. Thus, the device is selected when $V_{DD}$ (also used for MOSFET's gate-biasing in the decoder logic) approaches the threshold voltage ($V_{TH}$) of p-MOSFET. Note that n-MOSFET $V_{TH}$ is much lower due to the n++ doping of the Poly-Si gate. **(e)** Following the $V_{TH}$ tunability of n- and p-MOSFET (see also Supp. Fig. 2), the CMOS logic operation improves by making $V_{TH}$ of n- and p-MOSFET more symmetrical, thus allowing for CMOS logic to operate at $V_{DD}$ down to ~ 0.7 V. **(f-g)** Multiplexer cryo-CMOS logic leakage current measured between $V_{DD}$ and ground. The measured leakage is below the noise floor of the setup for both cases of $V_{BACK}$ = 0 V (f) and $V_{BACK}$ = −10 V (g). Thus, the quasi-dissipationless operation of our cryo-CMOS multiplexer is achieved.

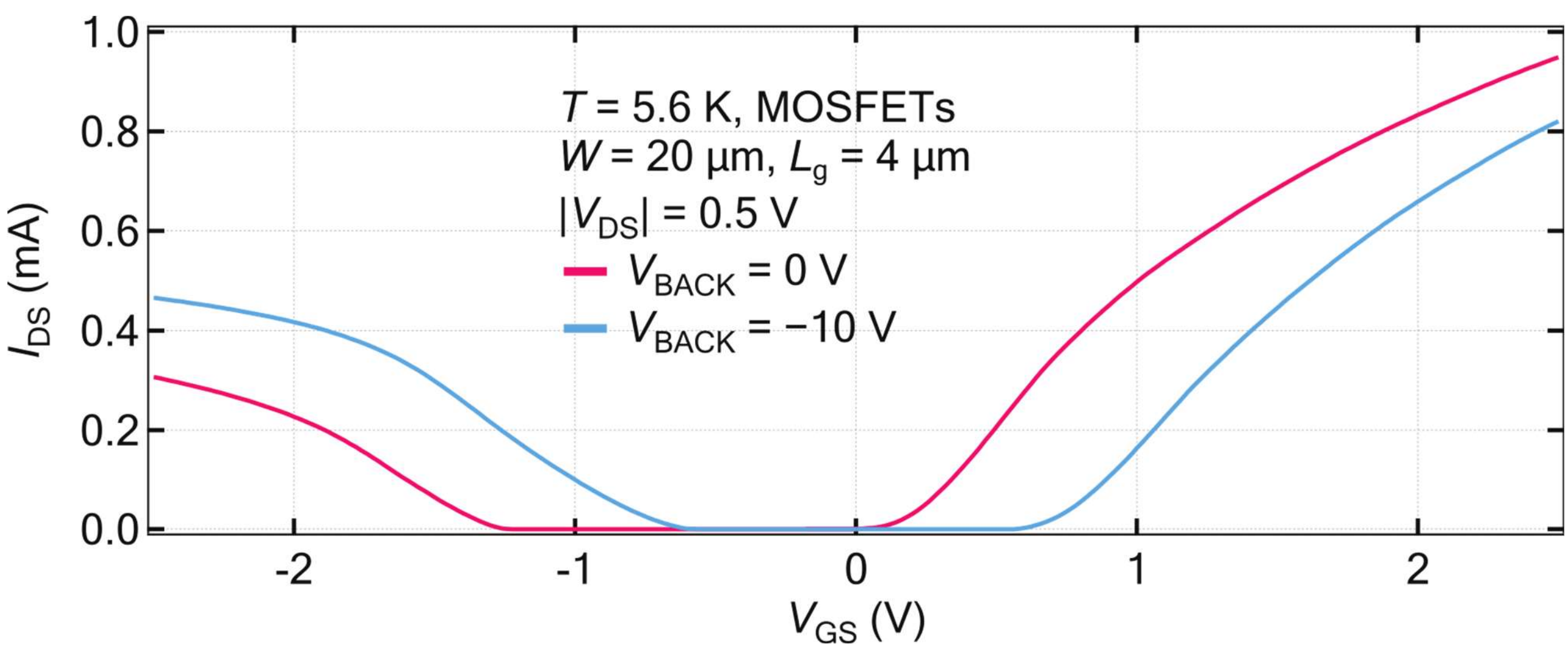

**Supplementary Fig. 2 | Back-gate tunability of cryo-CMOS MOSFETs.** The source-drain current dependence as a function of gate voltage $I_{DS}(V_{GS})$ is shown for n- and p-type metal oxide semiconductor field effect transistors (MOSFETs). Electron n- and hole p-MOSFETs have the same geometrical dimensions: channel width $W$ = 20 μm and gate length $L_g$ = 4 μm. Source-drain bias $V_{DS}$ = 0.5 V ($V_{DS}$ = −0.5 V) was used for n-MOSFET (p-MOSFET). Both kinds of transistors were measured at temperature $T$ = 5.6 K and two back-gate voltage values $V_{BACK}$ = 0 and −10 V. The threshold voltage $V_{TH}$ tunability with $V_{BACK}$ is demonstrated as follows. By applying more negative $V_{BACK}$ globally affecting both n- and p-MOSFET across the MUX chip, electron and hole transistor threshold voltages become more symmetric, thus, the cryo-CMOS supply and logic "1" voltages can be reduced as also discussed in Supp. Fig. 1.

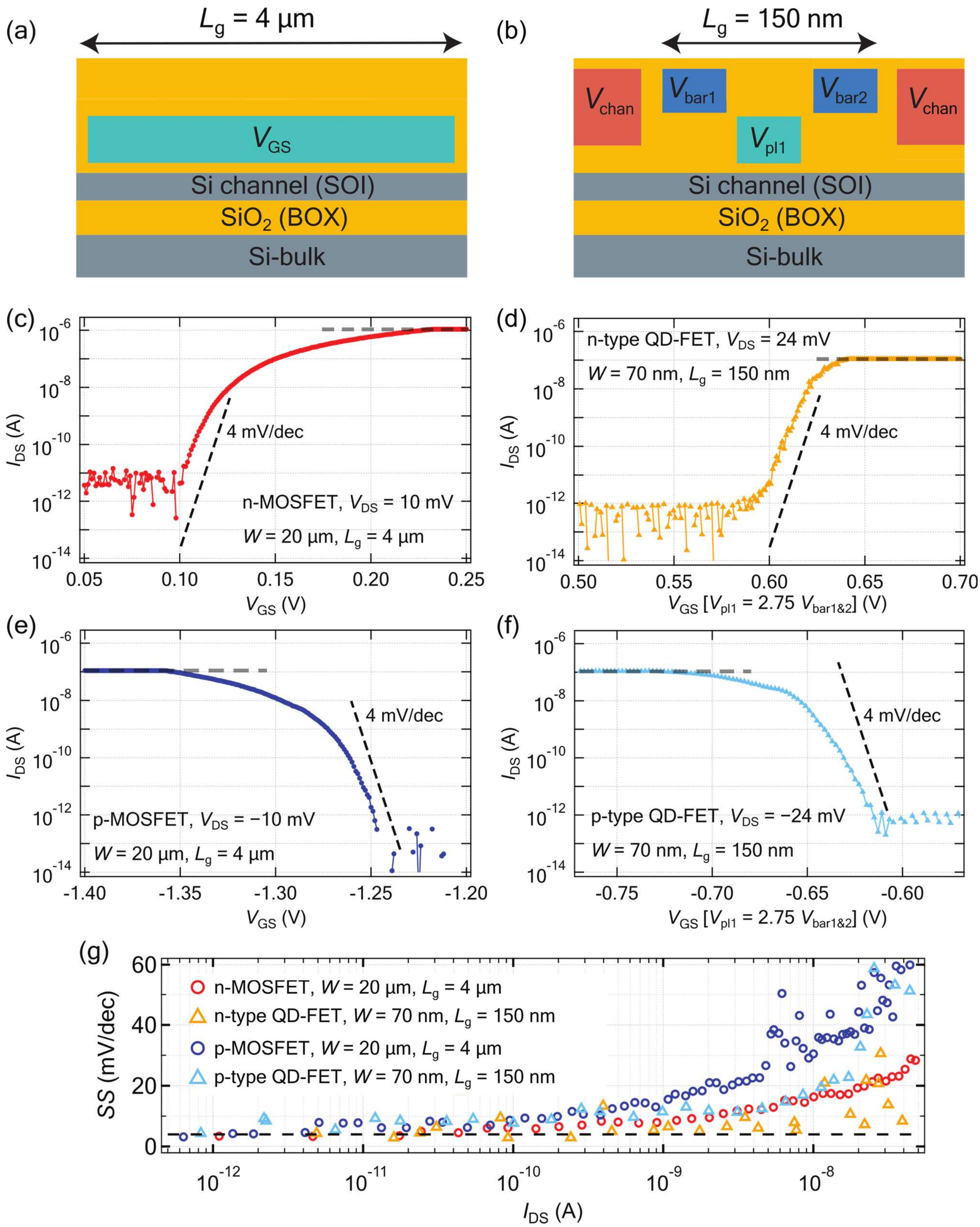

**Supplementary Fig. 3 | Transistor transfer characteristics and subthreshold swing. (a)** A schematic of the gate material stack of n- and p-type metal oxide semiconductor field effect transistors MOSFETs used in the cryo-CMOS multiplexers. **(b)** A schematic of the gate material stack of reconfigurable quantum dot–field effect transistor (QD-FET) devices, such as D[8], D[24], D[40], and D[56] for n-type, and D[9], D[25], D[41], and D[57] for p-type devices (see also Supp. table 1). In (a) and (b), the cross-section is taken along the channel direction, but for the MOSFET cross-section in

(a), the source and drain are not shown. In order to make a comparison between transfer characteristics (i.e., source-drain current dependence as a function of gate voltage, $I_{DS}(V_{GS})$) of MOSFETs and QD-FET devices, plunger $V_{pl1}$ and barrier gate voltages $V_{bar1}$ and $V_{bar2}$ of QD-FET devices are biased at the same time. Channel gate voltage $V_{chan}$ is used to create source and drain regions for QD-FET devices while for MOSFETs they are directly created by ion implantation phosphorous (n-type) and boron (p-type). However, to make gate-biasing on the gate-2 barrier gates equivalent to the gate-1 plunger gate and realize a QD-FET device with an effective 150 nm gate length (composed of three 50 nm gate lengths in series), $V_{bar1}$ and $V_{bar2}$ are biased with the voltage 2.75 times higher than the one applied to $V_{pl1}$. The factor of 2.75 comes from dividing 50 nm (gate-2 level $SiO_2$ oxide thickness) by 20 nm (gate-1 level $SiO_2$ oxide thickness). The $I_{DS}(V_{GS})$ curves shown in (c-f) were measured at back-gate voltage $V_{BACK} = 0$ V and temperature $T = 5.6$ K. **(c,e)** The $I_{DS}(V_{GS})$ characteristics of n- and p-type MOSFETs with gate length $L_g = 4$ μm used in the cryo-CMOS logic and analog switches at source-drain bias $|V_{DS}| = 10$ mV (linear source-drain biasing regime) are shown. **(d,f)** The $I_{DS}(V_{GS})$ characteristics of n- and p-type QD-FET (D[56] and D[57] from the multiplexer chip #1) with effective gate length $L_g = 150$ nm measured at $|V_{DS}| = 24$ mV (linear source-drain biasing regime) are shown. The reason for choosing $|V_{DS}| = 24$ mV for QD-FET devices was to apply a higher bias than the characteristic addition (charging) energy for electrons and holes in quantum dots as discussed in the main text. The horizontal gray dashed line in (c-f) indicates the saturation of the room temperature transimpedance amplifier. The tilted black dashed line in (c-f) indicates the subthreshold swing $SS$ with the slope of 4 mV/dec (as a guideline). Such low value of $SS$ is also achieved in the measurements of n- and p-type MOSFETs and QD-FETs. **(g)** The dependence of $SS$ as a function of $I_{DS}$ plotted for long-channel (n- and p-type MOSFETs) and short-channel (n- and p-type QD-FETs) transistors. The black dashed line indicates the ultra-low $SS$ value of 4 mV/dec achieved for all devices.

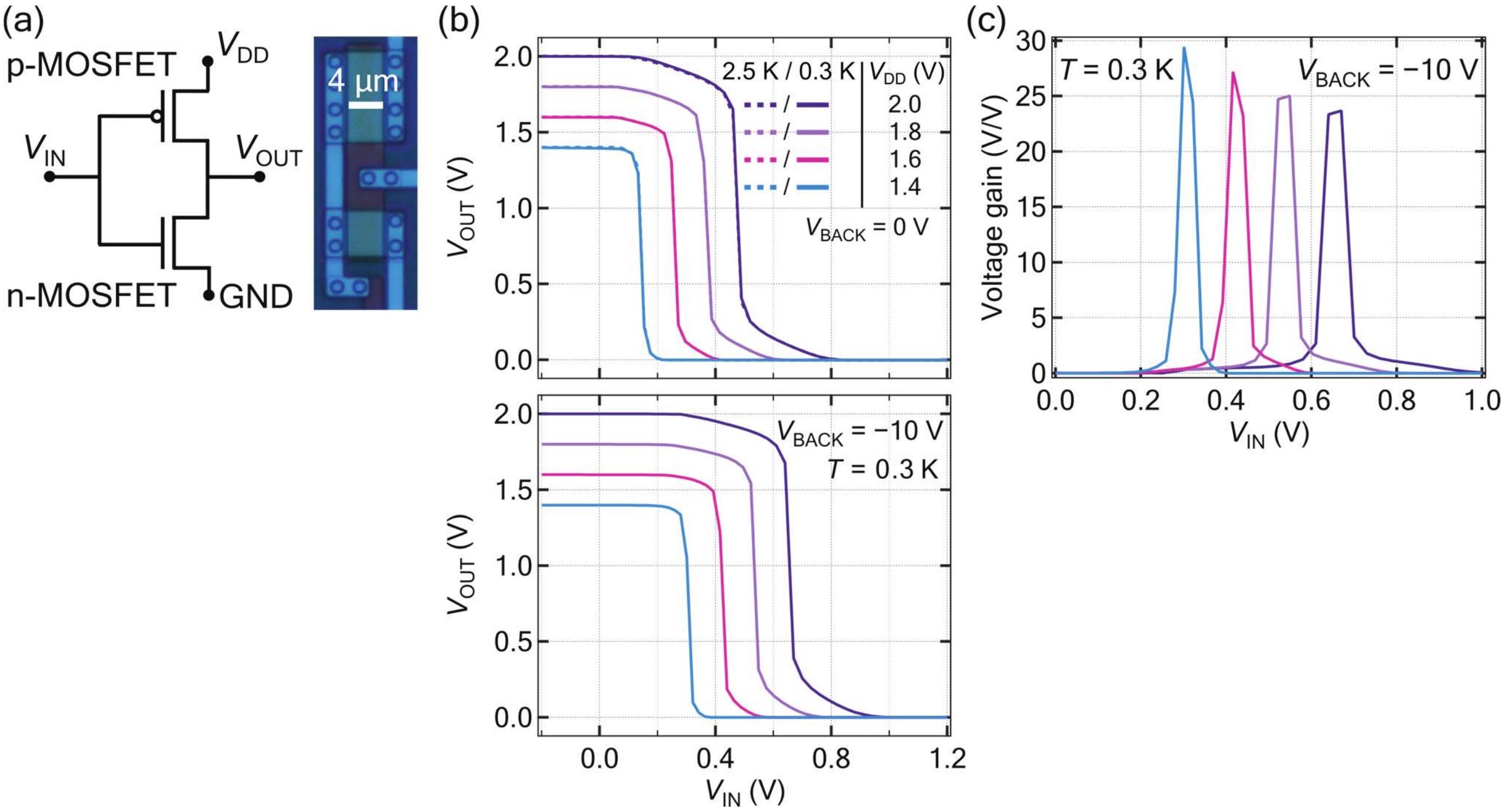

**Supplementary Fig. 4 | cryo-CMOS logic down to 300 mK temperature. (a)** A standard CMOS inverter (NOT-gate) topology is shown in the left panel. An optical micrograph of the fabricated inverter is shown in the right panel. Metal oxide semiconductor field effect transistors (MOSFETs) have gate length $L_g = 4$ μm and channel width $W$ of 5 μm for n-MOSFET and 10 μm for p-MOSFET. **(b)** In the top panel, the inverter characteristics of output voltage $V_{OUT}$ as a function

of input voltage $V_{IN}$ measured at the temperatures of $T = 2.5$ K and 300 mK are shown as a function of supply voltage $V_{DD}$ ranging from 1.4 to 2.0 V. One can observe that there is almost no difference between the $V_{OUT}(V_{IN})$ swing behavior of the CMOS inverter logic between 2.5 K and 300 mK, as expected from the cryogenic saturation of subthreshold swing and mobility. In the lower panel, the effect of negative back-gate biasing with $V_{BACK} = -10$ V on the inverter swing is shown. By making the threshold voltages $V_{TH}$ of electron n- and hole p-MOSFET more symmetrical, the $V_{OUT}(V_{IN})$ swing curves are shifted to the right along the horizontal axis of $V_{IN}$, thus improving the matching between transistors. The perfect matching between n-MOSFET and p-MOSFET in the inverter is achieved by choosing different channel widths for n- and p-type MOSFETs such that the swing is centered at $V_{IN} = V_{DD}/2$. **(c)** The inverter voltage gain $dV_{OUT}/dV_{IN}$ obtained by taking numerical derivatives of the data in the lower panel of (b). The sharp and narrow gain peak and the gain value reaching ~30 are typical features of commercial analog circuits based on advanced CMOS technologies.

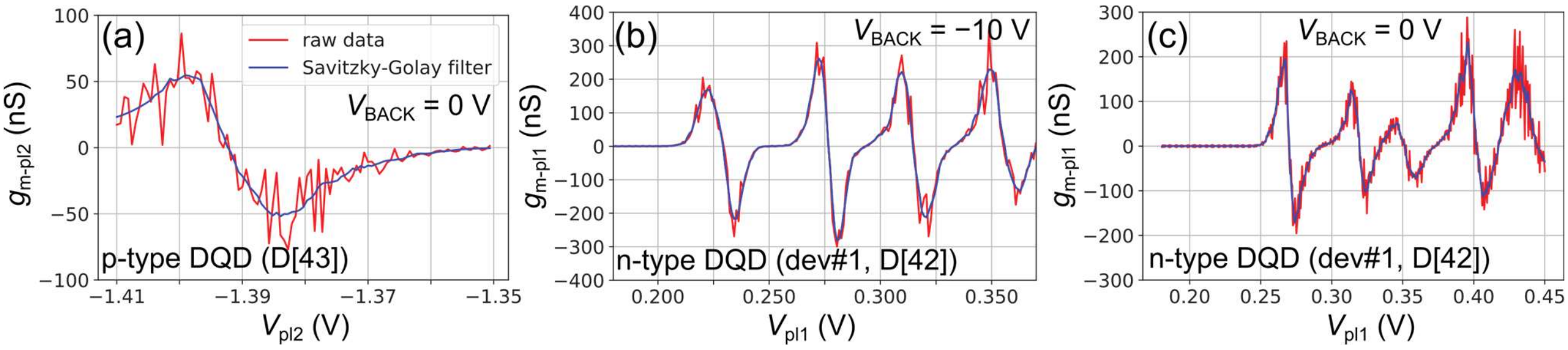

**Supplementary Fig. 5 | Numerical filtering of transconductance data for charge noise estimations.** Raw data for transconductance $g_{m\text{-}pl1(2)}$ from numerical derivatives of quantum dot current $I_{QD}$ with respect to the plunger gate voltage ($V_{pl1}$ and $V_{pl2}$ for controlling quantum dots QD1 and QD2 in double quantum dot devices, respectively) and numerically post-filtered data are shown for the following devices/quantum dots. **(a)** The hole quantum dot discussed in the main text with the measurement data shown in Supp. Fig. 6. **(b)** The electron quantum dot shown in Fig. 3. **(c)** The electron quantum dot shown in Supplementary Fig. 17 (same device and dot as in (c) but characterized at back-gate voltage $V_{BACK} = 0$ V).

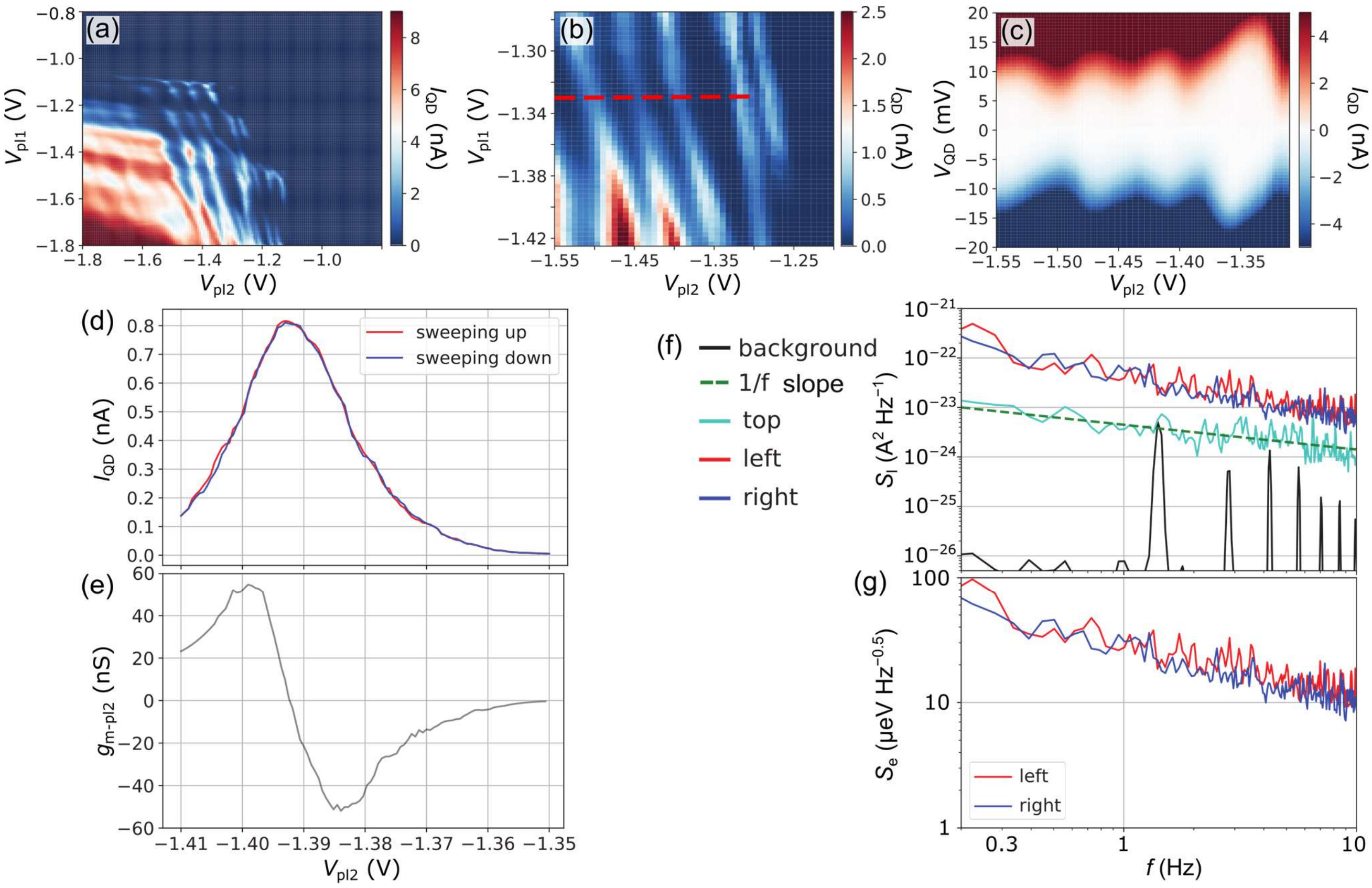

**Supplementary Fig. 6 | Hole double quantum dot device and hole charge noise. (a)** A stability diagram of a hole double quantum dot (DQD) device with the same geometry and gate layout as in the electron DQD devices discussed in Figs. 3-4 in the main text (dev#1, D[42] and dev#2, D[58]). The data in (a,b) are obtained by sweeping plunger gate voltages $V_{pl1}$ and $V_{pl2}$ and measuring source-drain current $I_{QD}$. Source-drain bias $V_{QD} = -3$ mV and channel gate voltage $V_{chan} = -3.7$ V were used. **(b)** A zoom over the stability diagram presented in (a). Two-dimensional Coulomb diamond measurements for quantum dot #2 (QD2) presented in **(c)** were performed at fixed $V_{pl1}$ and $V_{pl2}$ varying along the cut indicated with a red dashed line in (b). **(d)** The Coulomb blockade peak, which corresponds to the vertical cut of the dataset in (c) taken at $V_{QD}$ = 4 mV is shown (the absence of gate hysteresis is shown by sweeping up and down $V_{pl2}$ while measuring current). The lever arm parameter measured from the slopes of the Coulomb diamonds is 0.23 eV/V. **(e)** The numerical derivative of data in (d), or transconductance $g_{m\text{-}pl2}$, is shown. **(f)** The $S_I(f)$ curves are recorded at the $V_{pl2}$ points in the Coulomb blockade (background), on the top of a Coulomb diamond (top), and the right and left flanks of a Coulomb peak where $g_{m\text{-}pl2}$ reaches extremums (left and right). Additionally, the 1/f frequency dependence trend (1/f slope) is plotted. **(g)** Charge noise $S_e$ dependence as a function of $f$ estimated from data in (f) for the cases of $V_{pl2}$ tuned to the left and right flans of the Coulomb peak. The hole quantum dot $S_e$ value extracted at $f$ = 1 Hz is approximately 28 μeV $Hz^{-0.5}$.

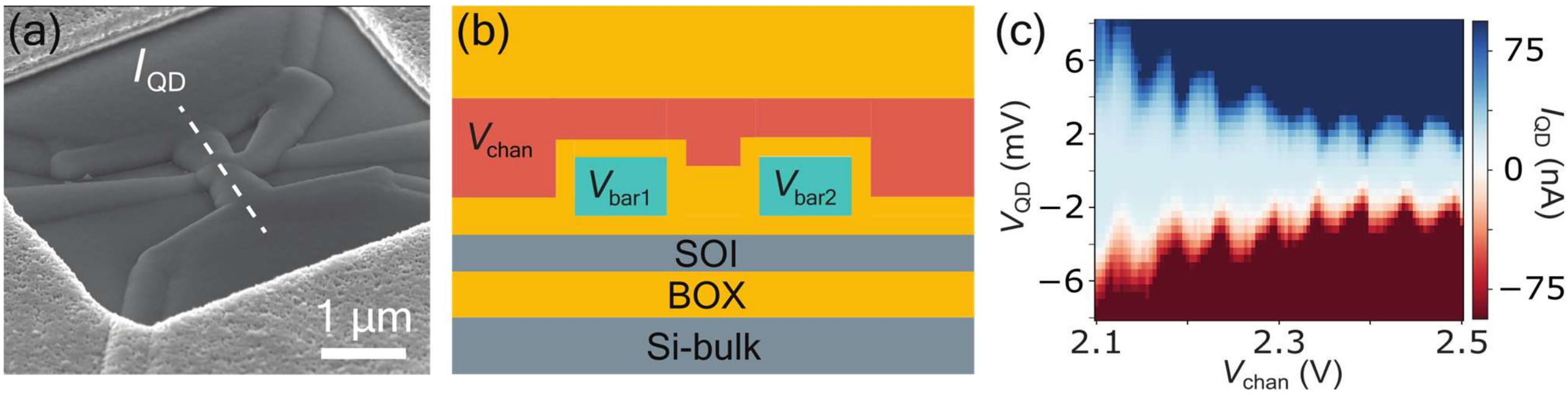

**Supplementary Fig. 7 | Electron quantum dot device measured at 300 mK.**

**(a)** A tilted scanning electron microscopy image of the test device measured with an 8-channel multiplexer (MUX) with the same circuit topology for the decoder and switches as in the case of 64-channel MUX chips #1 and #2 discussed in the main text and supplementary materials. **(b)** The cross-section schematic of the device shows the channel gate (gate-2) biased with $V_{chan}$ voltage, and two barrier gates (gate-1) biased with $V_{bar1}$ and $V_{bar2}$ voltages. **(c)** A coarse-resolution stability diagram showing two-dimensional Coulomb diamonds of the selected device measured at 300 mK. The measurement is obtained by fixing $V_{bar1}$ and $V_{bar2}$ and measuring quantum dot current $I_{QD}$ as a function of quantum dot voltage $V_{QD}$ and $V_{chan}$. Thus, it can be concluded that quantum dot devices and cryo-CMOS electronics, which both were studied at 5.6 K as discussed in the main text, also are functional at sub-Kelvin temperatures.

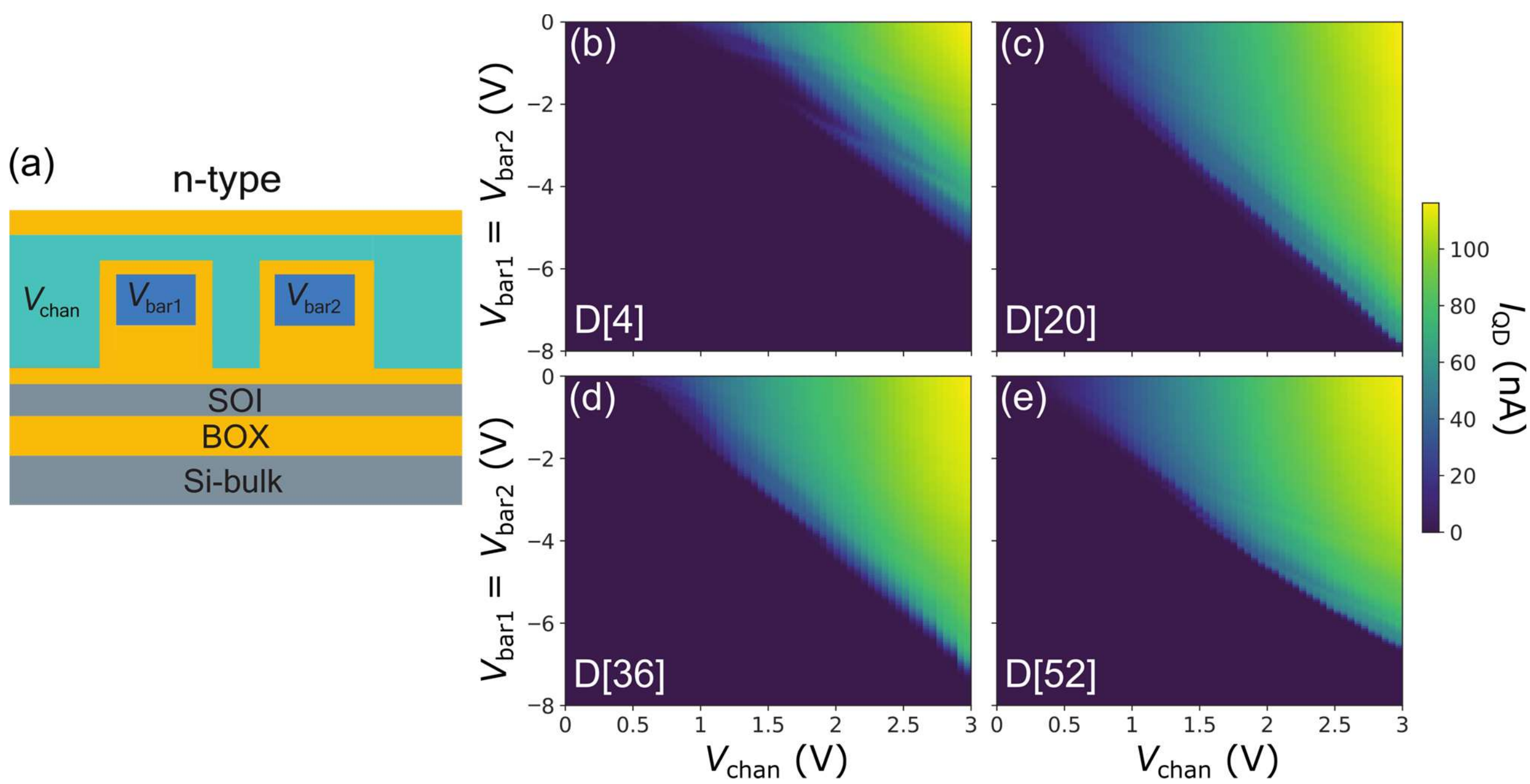

**Supplementary Fig. 8 | Characterization of electron quantum dot devices using the 64-channel MUX (chip #1).**

**(a)** A cross-section schematic of the electron (n-type) quantum dot device geometry. The stability diagrams in **(b-d)** were obtained by sweeping barrier gate (gate-2) voltages $V_{bar1}$ and $V_{bar2}$ at the same time vs channel gate (gate-1) voltage $V_{chan}$ and measuring quantum dot current $I_{QD}$. Supply voltage and address voltage $V_{DD}$ = $V_{add}$ = 1.5 V and quantum dot bias $V_{QD}$ = 3 mV were used. **(b-e)** Coarse-resolution stability diagrams for 4 nominally identical electron quantum dot devices D[4], D[20], D[36], and D[52] measured at temperature $T$ = 5.6 K in the 64-channel cryo-CMOS MUX, chip #1.

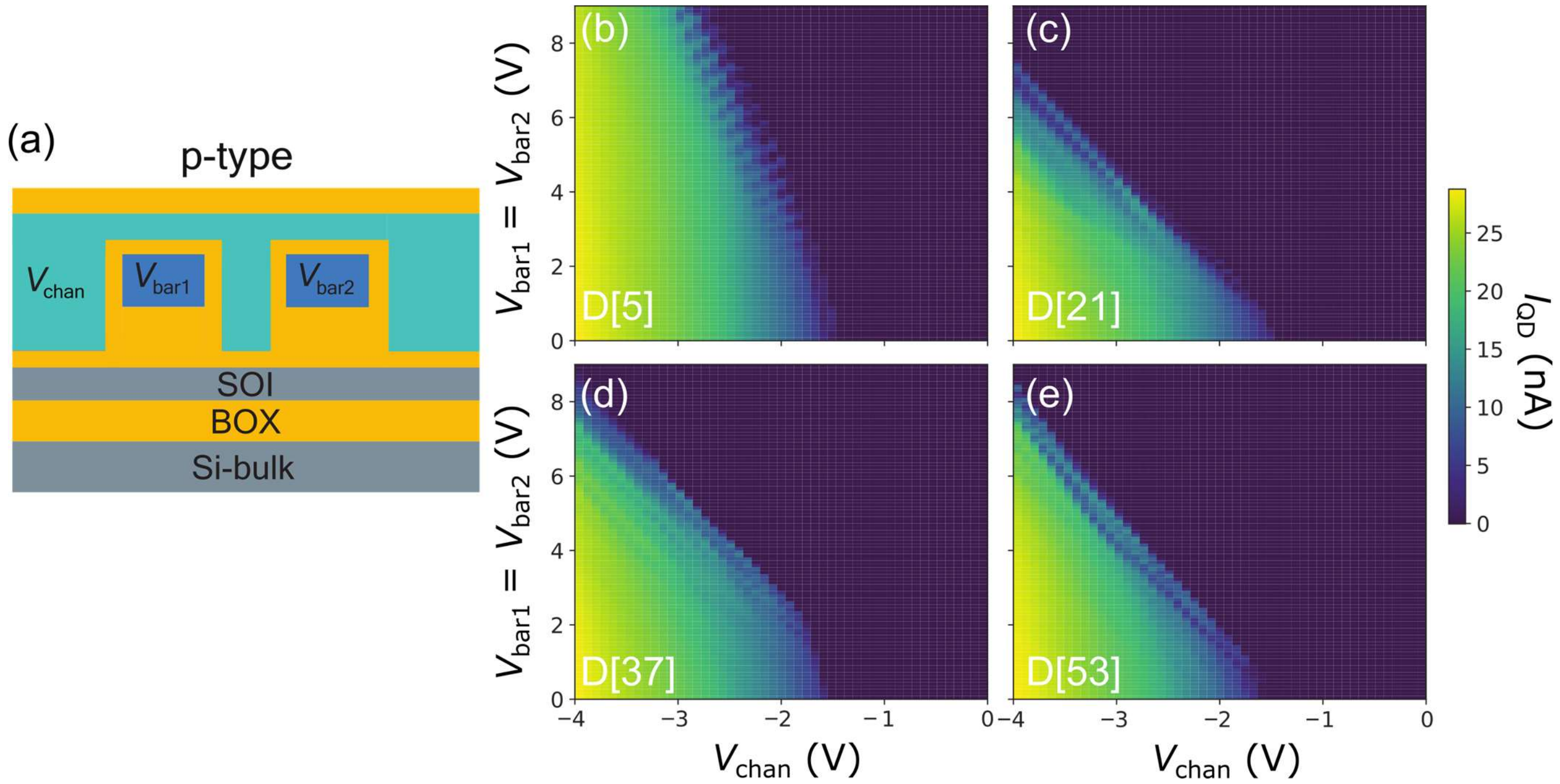

**Supplementary Fig. 9 | Characterization of hole quantum dot devices using the 64-channel MUX (chip #1).**

**(a)** A cross-section schematic of the hole (p-type) quantum dot device geometry. The stability diagrams in **(b-d)** were obtained by sweeping barrier gate (gate-2) voltages $V_{\mathrm{bar1}}$ and $V_{\mathrm{bar2}}$ at the same time vs channel gate (gate-1) voltage $V_{\mathrm{chan}}$ and measuring quantum dot current $I_{\mathrm{QD}}$. Supply voltage and address voltage $V_{\mathrm{DD}} = V_{\mathrm{add}} = 1.5$ V and quantum dot bias $V_{\mathrm{QD}} = -3$ mV were used. **(b-e)** Coarse-resolution stability diagrams for 4 nominally identical hole quantum dot devices D[5], D[21], D[37], and D[53] measured at temperature $T = 5.6$ K in the 64-channel cryo-CMOS MUX, chip #1.

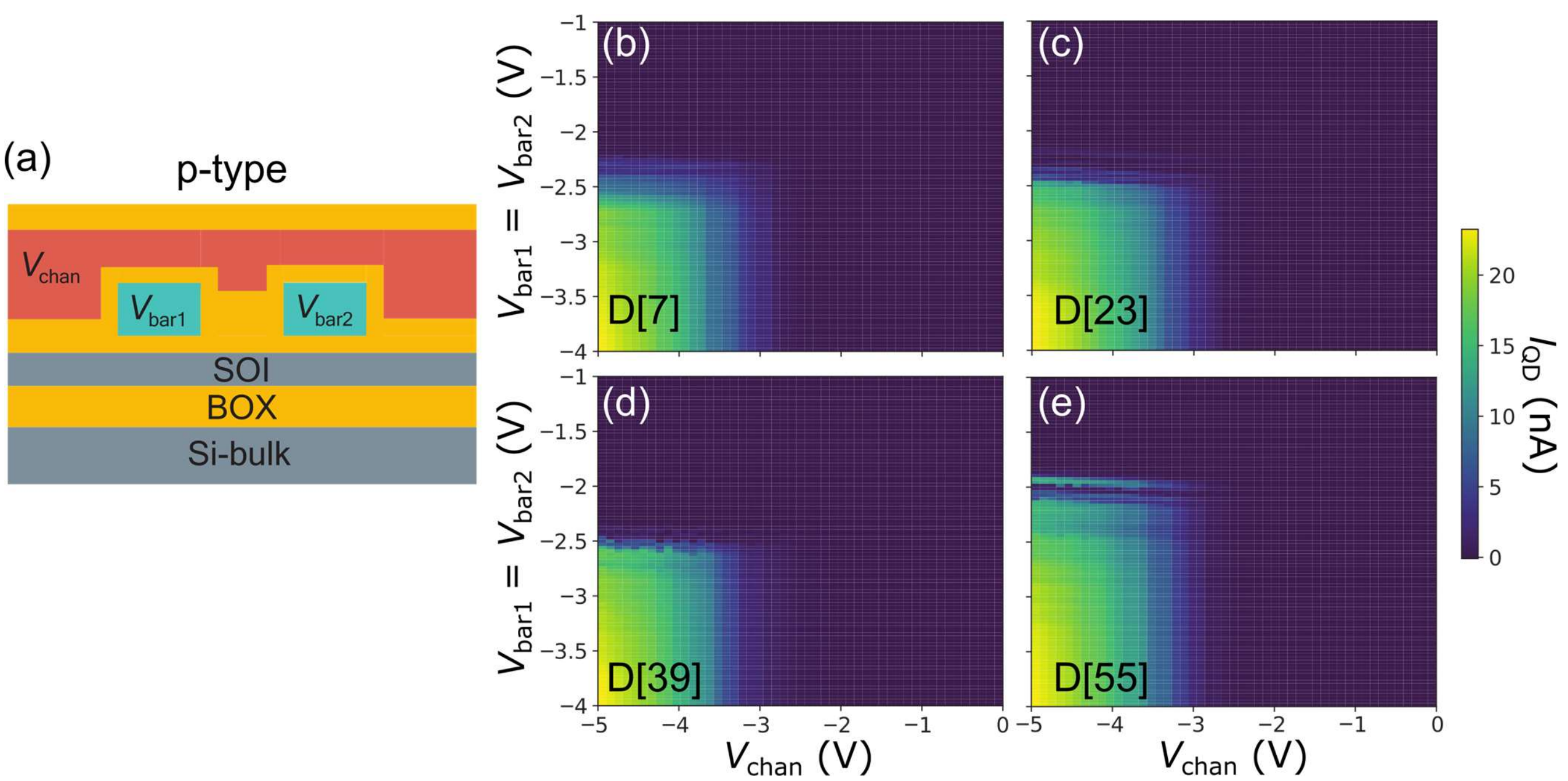

**Supplementary Fig. 10 | Characterization of hole quantum dot devices using the 64-channel MUX (chip #1).**

**(a)** A cross-section schematic of the hole (p-type) quantum dot device geometry. The stability diagrams in **(b-d)** were obtained by sweeping barrier gate (gate-1) voltages $V_{\mathrm{bar1}}$ and $V_{\mathrm{bar2}}$ at the same time vs channel gate (gate-2) voltage $V_{\mathrm{chan}}$ and measuring quantum dot current $I_{QD}$. Supply voltage and address voltage $V_{DD} = V_{add} = 1.5$ V and quantum dot bias $V_{QD} = -3$ mV were used. **(b-e)** Coarse-resolution stability diagrams for 4 nominally identical hole quantum dot devices D[7], D[23], D[39], and D[55] measured at temperature $T = 5.6$ K in the 64-channel cryo-CMOS MUX, chip #1.

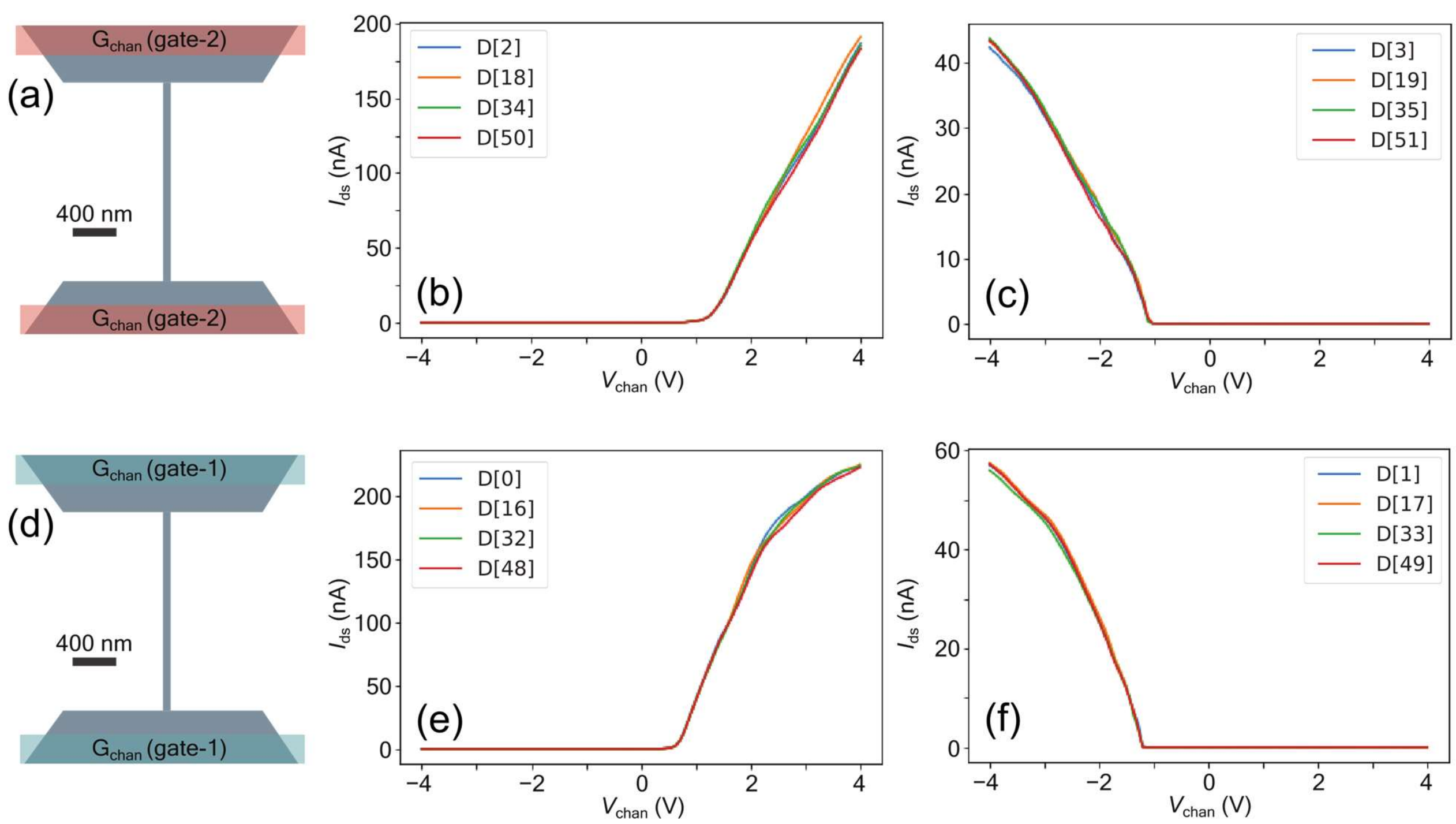

**Supplementary Fig. 11 | Characterization of nanowire devices using the 64-channel MUX (chip #2).**

**(a,d)** Top-view schematics of the electron (n-type) and hole (p-type) quantum dot device geometry with channel gate at gate-2 (a) and gate-1 (d) level. The field effect characteristics in **(b,c,e,f)** were obtained by sweeping channel gate voltage $V_{chan}$ and measuring quantum dot current $I_{QD}$. The same source-drain bias for each group of nominally identical devices was used. Supply voltage and address voltage $V_{DD} = V_{add} = 1.5$ V and back-gate voltage $V_{BACK} = 0$ mV were used. The measurements of the nanowire devices in the 64-channel cryo-CMOS MUX (chip #2) were done at temperature $T = 5.6$ K. Note the low variability in terms of the threshold voltage for n-type devices with G$_{chan}$ at gate-2 level (b), p-type devices with G$_{chan}$ at gate-2 level (c), n-type devices with G$_{chan}$ at gate-1 level (e), and p-type devices with G$_{chan}$ at gate-1 level (f).

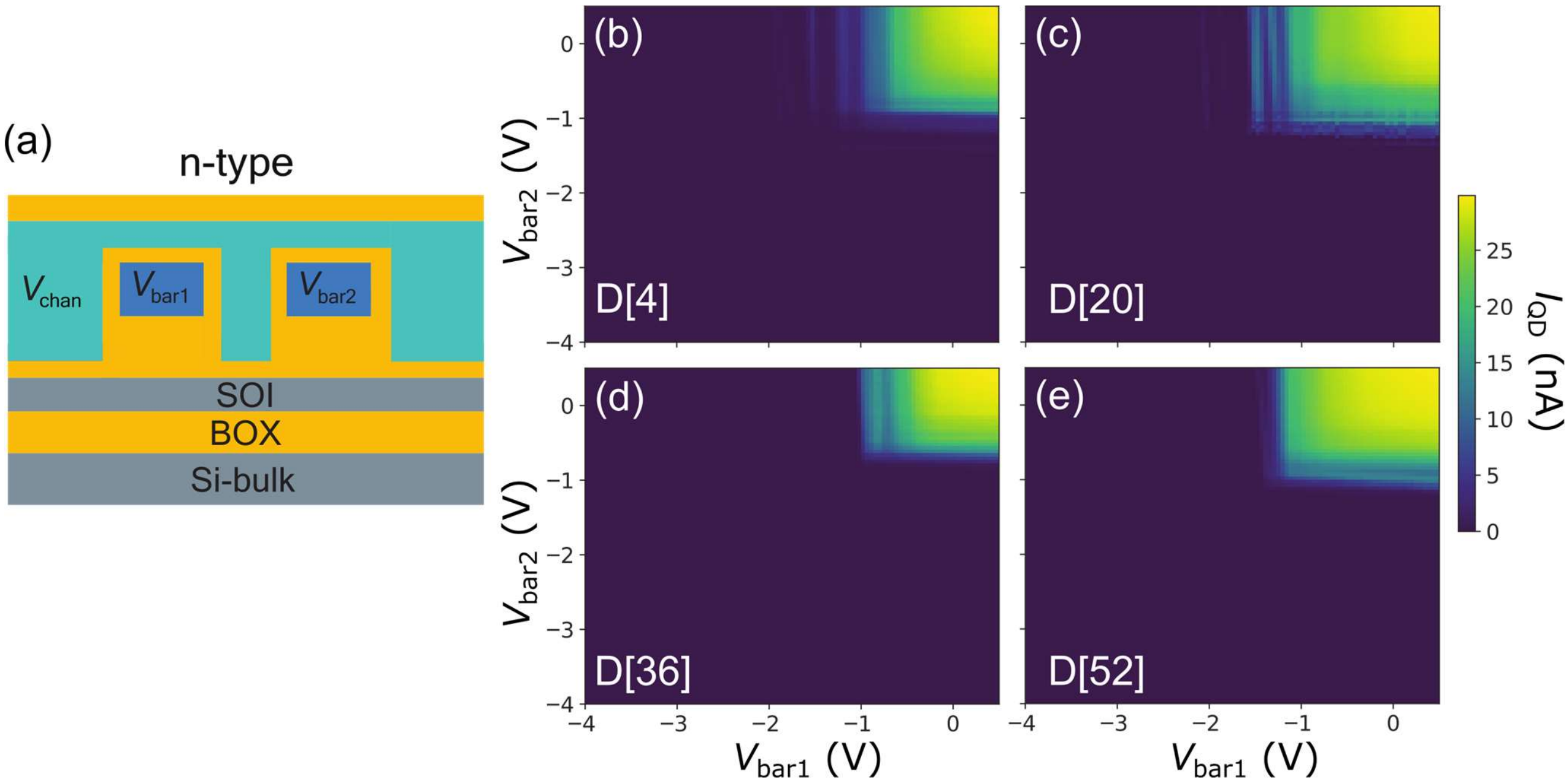

**Supplementary Fig. 12 | Characterization of electron quantum dot devices using the 64-channel MUX (chip #2).** **(a)** A cross-section schematic of the electron (n-type) quantum dot device geometry. The stability diagrams in **(b-d)** were obtained by sweeping barrier gate (gate-2) voltages $V_{bar1}$ and $V_{bar2}$ and measuring quantum dot current $I_{QD}$ with channel gate (gate-1) voltage $V_{chan}$ fixed at 1 V. Supply voltage and address voltage $V_{DD} = V_{add} = 1.5$ V and quantum dot bias $V_{QD}$ = 2 mV were used. **(b-e)** Coarse-resolution stability diagrams for 4 nominally identical electron quantum dot devices D[4], D[20], D[36], and D[52] measured at temperature $T = 5.6$ K in the 64-channel cryo-CMOS MUX, chip #2.

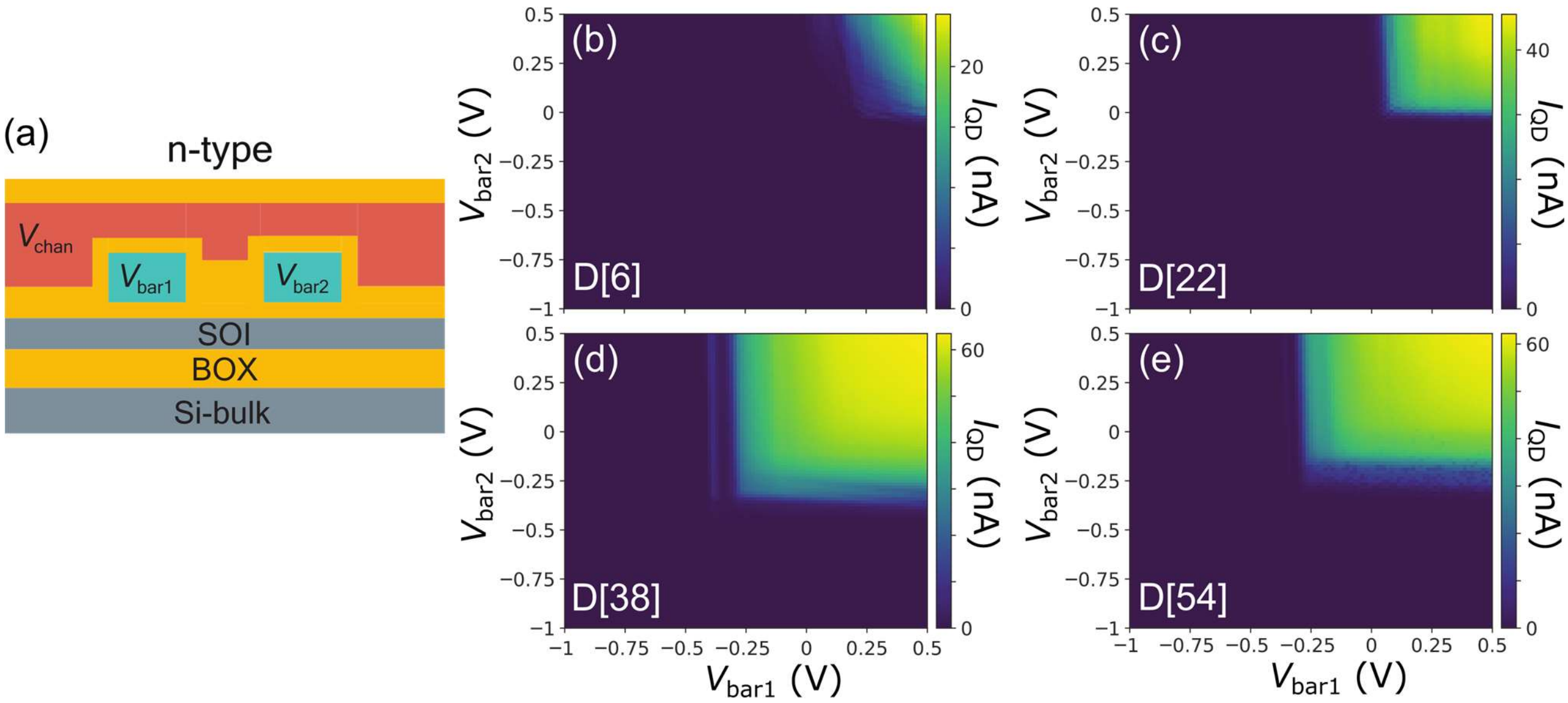

**Supplementary Fig. 13 | Characterization of electron quantum dot devices using the 64-channel MUX (chip #2).** **(a)** A cross-section schematic of the electron (n-type) quantum dot device geometry. The stability diagrams in **(b-d)** were obtained by sweeping barrier gate (gate-1) voltages $V_{bar1}$ and $V_{bar2}$ and measuring quantum dot current $I_{QD}$ with channel gate (gate-2) voltage $V_{chan}$ fixed at 3.5 V. Supply voltage and address voltage $V_{DD} = V_{add} = 1.5$ V and quantum dot bias $V_{QD}$

= 2 mV were used. **(b-e)** Coarse-resolution stability diagrams for 4 nominally identical electron quantum dot devices D[6], D[22], D[38], and D[54] measured at temperature $T = 5.6$ K in the 64-channel cryo-CMOS MUX, chip #2.

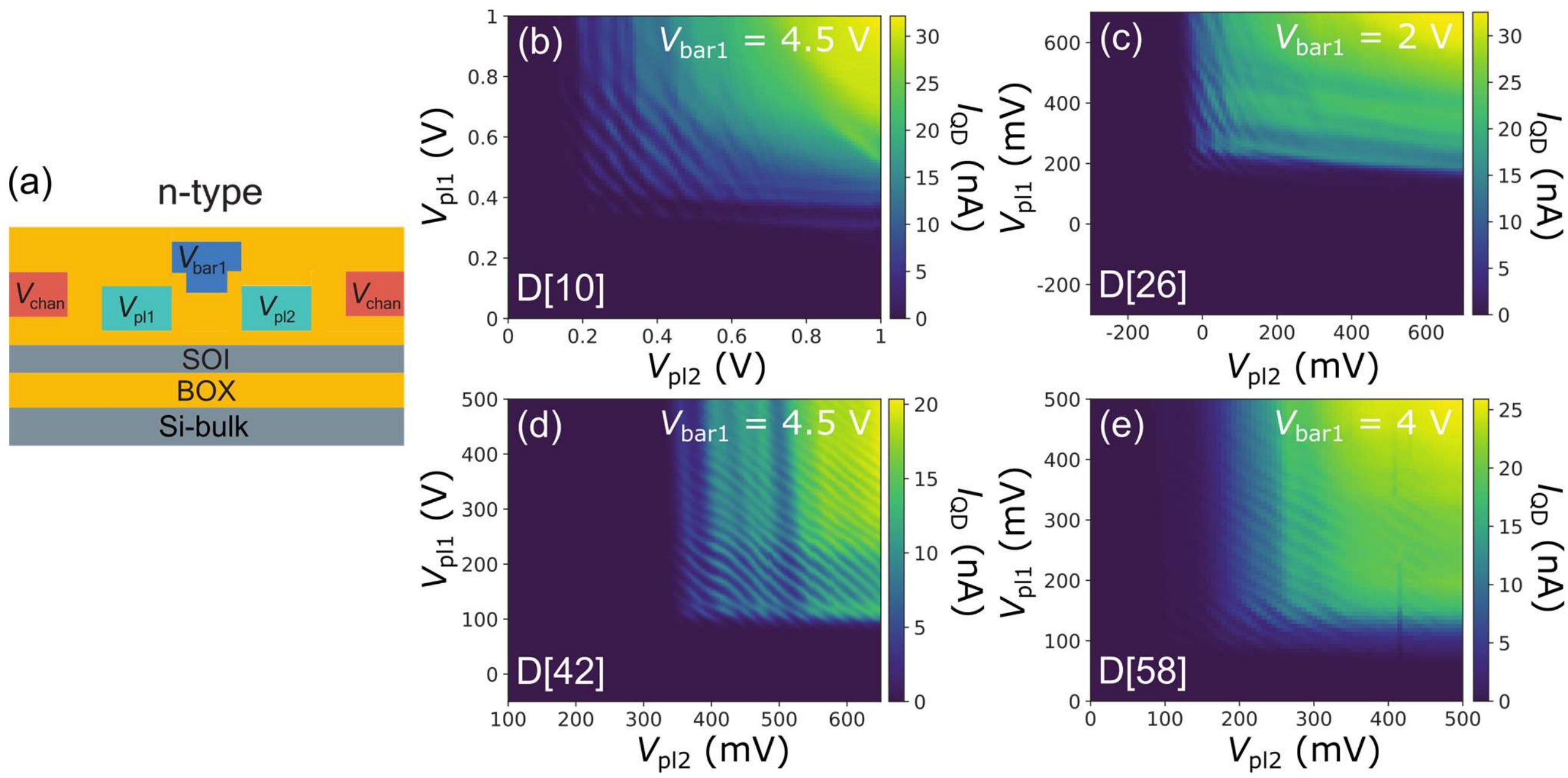

**Supplementary Fig. 14 | Characterization of electron double quantum dots using the 64-channel MUX (chip #1). (a)** A cross-section schematic of the electron (n-type) double quantum dot device geometry. The stability diagrams in **(b-d)** were obtained by sweeping plunger gate (gate-1) voltages $V_{pl1}$ and $V_{pl2}$ and measuring quantum dot current $I_{QD}$ with channel gate (gate-2) voltage $V_{chan}$ fixed at 3 V and fixed barrier gate (gate-2) voltage $V_{bar1}$. Supply voltage and address voltage $V_{DD} = V_{add} = 1.5$ V and quantum dot bias $V_{QD} = 2$ mV were used. **(b-e)** Fine-tuned in the strong coupling regime with $V_{bar1}$, stability diagrams for 4 nominally identical electron double quantum dot devices D[10], D[26], D[42], and D[58] measured at temperature $T = 5.6$ K and back-gate voltage $V_{BACK} = -10$ V in the 64-channel cryo-CMOS MUX, chip #1. The devices D[42] and D[58] are analyzed in detail in the main text and Fig. 3.

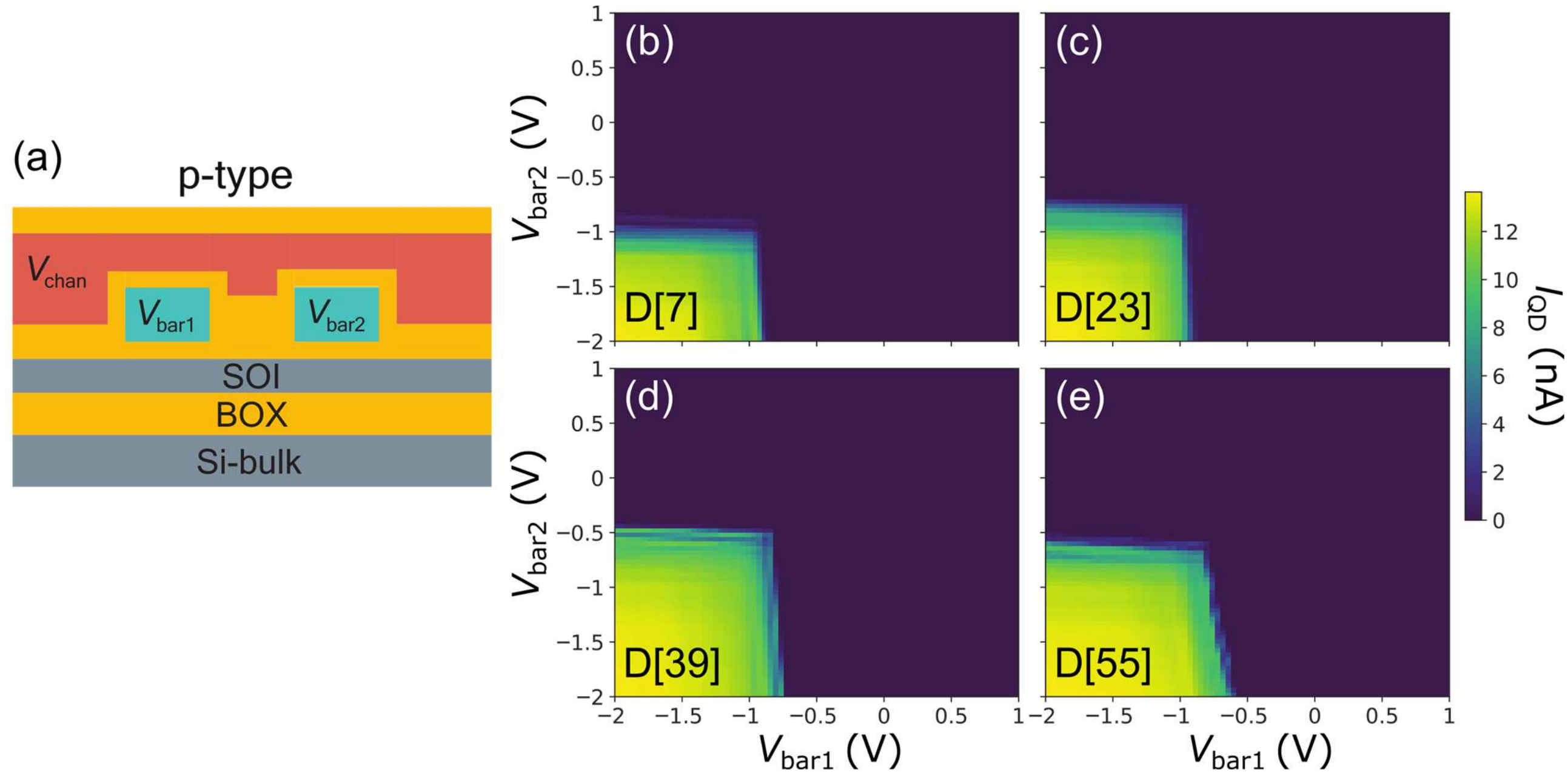

**Supplementary Fig. 15 | Characterization of hole quantum dot devices using the 64-channel MUX (chip #2).**

**(a)** A cross-section schematic of the hole (p-type) quantum dot device geometry. The stability diagrams in **(b-d)** were obtained by sweeping barrier gate (gate-1) voltages $V_{\text{bar1}}$ and $V_{\text{bar2}}$ and measuring quantum dot current $I_{\text{QD}}$ with channel gate (gate-2) voltage $V_{\text{chan}}$ fixed at −4.5 V. Supply voltage and address voltage $V_{\text{DD}} = V_{\text{add}} = 1.5$ V and quantum dot bias $V_{\text{QD}} = -2$ mV were used. **(b-e)** Coarse-resolution stability diagrams for 4 nominally identical hole quantum dot devices D[7], D[23], D[39], and D[55] measured at temperature $T = 5.6$ K in the 64-channel cryo-CMOS MUX, chip #2.

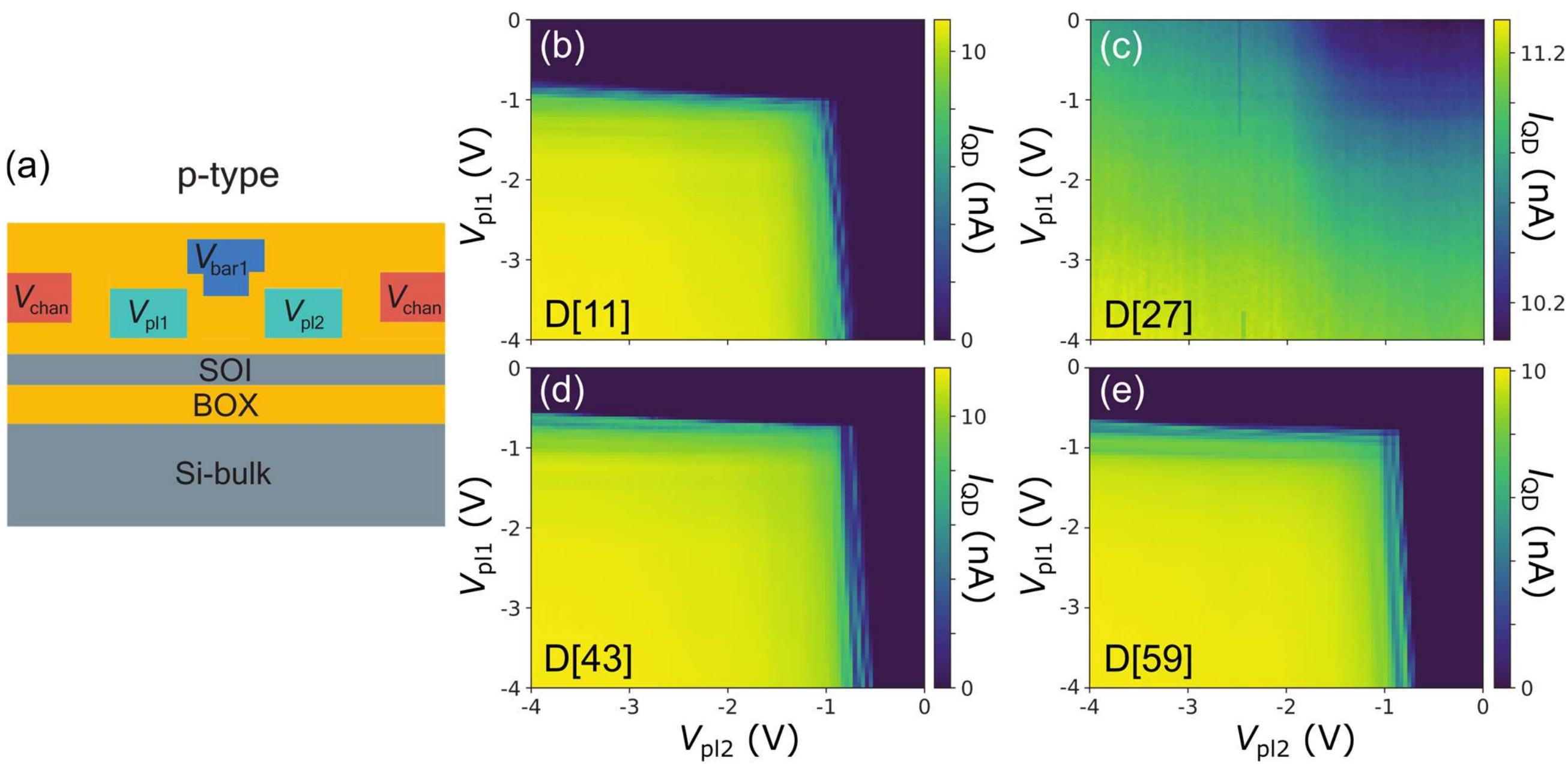

**Supplementary Fig. 16 | Characterization of hole quantum dot devices using the 64-channel MUX (chip #1).**

**(a)** A cross-section schematic of the hole (p-type) double quantum dot device geometry. The stability diagrams in **(b-d)** were obtained by sweeping plunger gate (gate-1) voltages $V_{\text{pl1}}$ and $V_{\text{pl2}}$ and measuring quantum dot current $I_{\text{QD}}$ with channel gate (gate-2) voltage $V_{\text{chan}}$ fixed at − 3 V and barrier gate (gate-2) voltage $V_{\text{bar1}} = -6$ V. Supply voltage and address voltage $V_{\text{DD}} = V_{\text{add}} = 1.5$ V and quantum dot bias $V_{\text{QD}} = -2$ mV were used. **(b-e)** Stability diagrams for 4 nominally identical hole double quantum dot devices D[11], D[27], D[43], and D[59] measured at temperature $T$ = 5.6 K and back-gate voltage $V_{BACK}$ = −10 V in the 64-channel cryo-CMOS MUX, chip #1. The device D[43] was tuned up to characterized quantum dot parameters and low-frequency charge noise, see the discussion in the main text and fine-tuned stability diagram and Coulomb blockade data presented in Supp. Fig. 6.

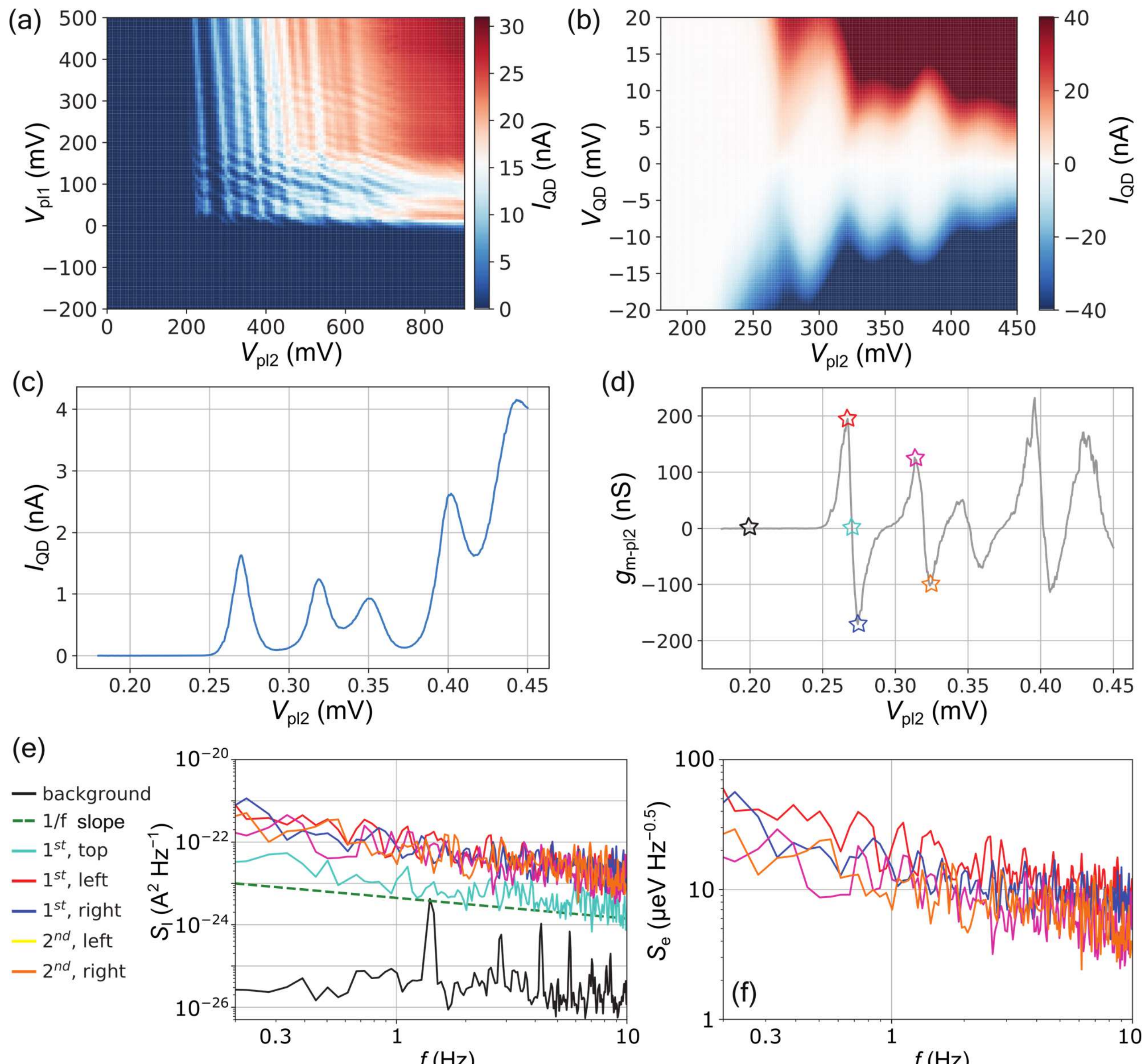

**Supplementary Fig. 17 | Electron double quantum dot device (dev#1 from the main text) at $V_{BACK}$ = 0V. (a)** The stability diagram of the electron double quantum dot (DQD) dev#1 (D[42]) measured by sweeping plunger gate voltages $V_{pl1}$ and $V_{pl2}$ and measuring quantum dot current $I_{QD}$ at back-gate voltage $V_{BACK}$ = 0 V is shown. The data was acquired at temperature $T$ = 5.6 K and barrier gate voltage $V_{bar1}$ = 3.5 V. **(b)** Two-dimensional Coulomb diamonds for quantum dot QD2 measured in the regime of weakly coupled quantum dots set by $V_{bar1}$ = 2 V are shown. **(c)** Horizontal cut of data in (b) at $V_{QD}$ = 1 mV. **(d)** The numerical derivative of data from (c) is plotted as transconductance $g_{m\text{-}pl2}$ as a function of $V_{pl2}$. **(e)** Low-frequency current noise measurements for several $V_{pl2}$ points according to the start markers in (d). The $S_I(f)$ curves are recorded at the $V_{pl2}$ points in the Coulomb blockade (background), on the top of the first Coulomb diamond ($1^{st}$, top),

and the right and left flanks of the first two detected Coulomb peaks (1st and 2nd, left and right). Additionally, the 1/f frequency dependence trend (1/f slope) is plotted. **(f)** Charge noise dependence as a function of frequency $S_e(f)$ calculated based on data in (e) is plotted. Note that, unlike the current noise data shown in the main text which were averaged for several minutes each, here the noise was measured in a single-shot manner, so the measurements appear noisier as compared to data shown in Fig. 4 of the main text. Nevertheless, very similar values of charge noise for the cases of $V_{BACK}$ = 0V presented here and $V_{BACK}$ = −10 V studied in Fig. 4 of the main text are obtained.

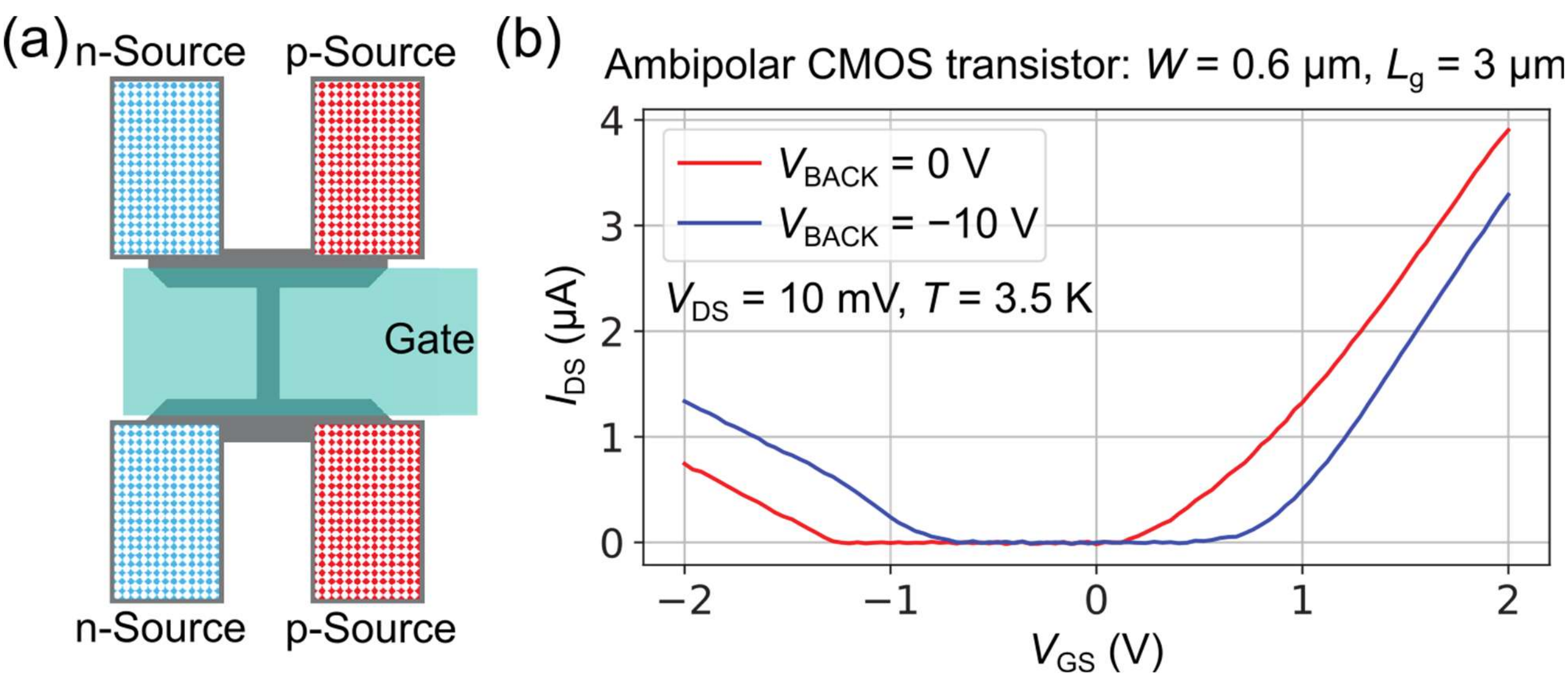

**Supplementary Fig. 18 | Ambipolar CMOS transistors. (a)** Sketch showing the ambipolar CMOS transistors fabricated on the same wafer as the 64-channel multiplexer chips. In such ambipolar devices, the electron or hole channel is accumulated depending on gate-voltage polarization. The metal oxide semiconductor field effect transistors (MOSFETs) feature two pairs of electron and hole Source and Drain connected to each end of the device. **(b)** Transfer characteristics of source-drain current $I_{DS}$ as a function of gate voltage $V_{GS}$ of the ambipolar transistors measured at $V_{BACK}$ = 0 V and $V_{BACK}$ = − 10 V at temperature $T$ = 3.5 K and source-drain bias $V_{DS}$ = 10 mV. The transistor shaped as a nanowire has the channel width $W$ = 0.6 μm and gate length $L_g$ = 3 μm.